\documentclass[12pt]{article}

\usepackage[letterpaper,margin=1in]{geometry}
\usepackage{blindtext}
\usepackage{graphicx}
\usepackage{amsthm}
\usepackage{amssymb}
\usepackage[authoryear,round]{natbib}

\usepackage{amsmath}
\DeclareMathOperator*{\argmin}{argmin}
\usepackage{hyperref}
\usepackage{amsfonts}
\usepackage{enumerate}
\usepackage{url} 
\usepackage{color}
\usepackage{multicol}
\usepackage{multirow}
\usepackage{booktabs}
\usepackage{makecell}
\usepackage{tikz}
\usepackage{algorithm}
\usepackage{algpseudocode}
\usetikzlibrary{arrows.meta,positioning,fit,calc}
\usepackage{color,soul}

\newcommand{\bphi}{ \boldsymbol{\phi} }
\newcommand{\bpsi}{ \boldsymbol{\psi} }
\newcommand{\bzeta}{ \boldsymbol{\zeta} }

\newcommand{\btheta}{\boldsymbol{\theta}}

\newcommand{\bTheta}{ \boldsymbol{\Theta} }

\newcommand{\bgamma}{ \boldsymbol{\gamma} }

\newcommand{\bdelta}{ \boldsymbol{\delta} }

\newcommand{\bzero}{\textbf{0}}

\newcommand{\bZ}{\textbf{Z}}
\newcommand{\bz}{\textbf{z}}

\newcommand{\bb}{\textbf{b}}

\newcommand{\bc}{\textbf{c}}

\newcommand{\be}{\textbf{e}}
\newcommand{\bE}{\textbf{E}}

\newcommand{\bh}{\textbf{h}}
\newcommand{\bH}{\textbf{H}}

\newcommand{\bg}{\textbf{g}}

\newcommand{\bs}{\textbf{s}}

\newcommand{\bU}{\textbf{U}}

\newcommand{\bx}{\boldsymbol{x}}

\newcommand{\by}{\boldsymbol{y}}

\newtheorem{theorem}{Theorem}
\newtheorem{proposition}[theorem]{Proposition}
\newtheorem{corollary}[theorem]{Corollary}
\theoremstyle{remark}
\newtheorem{remark}[theorem]{Remark}
\newcommand{\BlackBox}{\ensuremath{\blacksquare}}

\def\spacingset#1{\renewcommand{\baselinestretch}{#1}\small\normalsize}
\spacingset{1}

\begin{document}

\title{\bf Divide-and-Conquer Neural Bayes Estimation for Spatial and Spatiotemporal Point Process Models}

\author{Bokgyeong Kang$^{1}$, Andrew Zammit-Mangion$^{2}$,\\
Erin M. Schliep$^{3}$, and Robert S. Schick$^{4}$\\[0.5em]
\small $^{1}$Department of Statistics, Dongguk University, Seoul, Republic of Korea\\
\small $^{2}$School of Mathematics and Statistics, University of New South Wales, Sydney, Australia\\
\small $^{3}$Department of Statistics, North Carolina State University, Raleigh, USA\\
\small $^{4}$Southall Environmental Associates, Aptos, USA}
\date{}

\maketitle

\begin{abstract}
Spatial and spatiotemporal point process models are widely used to analyze event data across many scientific disciplines, but likelihood-based Bayesian inference is often computationally prohibitive due to intractable likelihoods, high-dimensional latent structures, and dependence on event history. Existing approximate and likelihood-free methods mitigate some of these challenges, but often at the cost of reduced statistical efficiency, imperfect uncertainty quantification, or information loss from reliance on predefined summary statistics. A promising recent approach involves neural networks, but it has only been explored in purely spatial or temporal settings and requires a computationally expensive training stage for large domains or dense point patterns. Here, we propose a divide-and-conquer neural Bayes estimation framework for spatial and spatiotemporal point process models. Our method operates directly on raw point pattern realizations using transformer-based architectures on partitions of the spatiotemporal domain, together with an attention mechanism to optimally combine locally extracted summaries. Simulation studies demonstrate reliable inference with improved scalability relative to existing general-purpose neural inference approaches and classical likelihood-based methods. We further demonstrate the practical applicability of our approach through an analysis of whale call data in Cape Cod Bay, MA, where, once trained, the estimator yields parameter estimates and credible intervals at negligible computational cost.
\end{abstract}


\newpage
\spacingset{1.45}

\section{Introduction}
\label{sec:intro}

Spatial and spatiotemporal point pattern data are routinely observed across a wide range of scientific disciplines, including ecology, epidemiology, seismology, neuroscience, and finance \citep{wikle2001spatiotemporal,hooten2008hierarchical,moller2010structured,Clements2011,Meyer2012,Kang2025whaleHawkes}.
In these applications, data are often modeled as realizations of a point process, where interest lies in understanding the underlying mechanisms governing event occurrence, interaction, and clustering in time and space.
A rich class of point process models has been developed for this purpose, which includes Poisson and Cox processes for spatial heterogeneity \citep{kingman1992poisson}; log-Gaussian Cox processes (LGCPs) for latent spatial structure \citep{moller1998log}; Strauss processes for pairwise spatial interaction and inhibition \citep{strauss1975model,kelly1976note}; and Hawkes-type processes for history-dependent self-excitation and interaction dynamics in temporal or spatiotemporal settings \citep{Hawkes1971}.
These models provide interpretable statistical descriptions of complex dependence structures and have been widely used to address substantive social and scientific questions \citep[e.g.,][]{zammit2012point}.
Despite their modeling flexibility, making inference with spatial and spatiotemporal point process models is computationally demanding.
Likelihood-based approaches have traditionally relied on Markov chain Monte Carlo (MCMC) or Hamiltonian Monte Carlo (HMC) methods \citep[e.g.,][]{adams2009tractable}.
For most point process models, the likelihood is expensive to evaluate.
Computational hurdles stem from required intractable integral evaluations over large spatial or spatiotemporal domains, high-dimensional latent structures, or dependence on the event history. This intractability leads to computational bottlenecks when analyzing large, heterogeneous point patterns.

To mitigate the computational burden, a variety of approximate inference strategies  for point process models have been proposed.
A strand of the literature focuses on scalable approximations to latent Gaussian processes, including nearest-neighbor Gaussian processes \citep{shirota2019scalable}, Vecchia-type approximations \citep{zilber2021vecchia}, and low-rank representations based on low-dimensional basis functions \citep{zammit2012variational}, which reduce computational cost by exploiting conditional independence or low-rank structure at the expense of potential loss in statistical efficiency. A second class of approaches relies on likelihood or posterior approximations, such as composite likelihood methods \citep{guan2006composite} and, specifically, Palm likelihoods \citep{Collins2025}, integrated nested Laplace approximation (INLA) \citep{Rue2009,flagg2023integrated}, and variational Bayes methods \citep{lloyd2015variational,cseke2016sparse}. These methods offer substantial computational gains, but often require model-specific derivations, and sometimes misrepresent posterior uncertainty. When likelihood evaluation is infeasible, likelihood-free inference methods such as approximate Bayesian computation \citep[ABC;][]{shirota2017approximate,vihrs2022approximate} offer an alternative. These bypass explicit likelihood computation, but are typically sensitive to the choice of summary statistics and may lead to information loss.

More recently, neural network-based methods have been proposed for parameter estimation that, like ABC, do not require explicit likelihood evaluation.
These approaches involve a neural network trained on simulated data to learn a mapping from data realizations to model parameters. These inferential methods are said to be \emph{amortized} since they typically require a computationally intensive training stage but, once trained, can be used for rapid inference on different datasets.
In spatial statistics, neural estimation has primarily been developed for geostatistical (point-referenced) data. 
For example, \citet{gerber2021fast} train a neural network to predict covariance function parameters in Gaussian process models, while \citet{lenzi2023neural} use a neural network for parameter estimation in max-stable process models for spatial extremes.
\citet{Sainsbury-Dale2024} construct ``neural Bayes'' estimators to approximate Bayes estimators, and they employ permutation-invariant neural networks to make inference from replicated gridded spatial data. This framework is subsequently extended by \citet{Sainsbury-Dale2025} to irregular spatial settings through the use of graph neural networks. In a related direction, \citet{Hector2024} propose a black-box divide-and-conquer framework that partitions the spatial or spatiotemporal domain into blocks, performs neural estimation within each block, and combines the resulting estimates via bootstrapping and weighted averaging.

For point process models, neural network-based estimation has been developed in both temporal and spatial settings.
In the temporal setting, \citet{Lee2023} use long short-term memory (LSTM) networks for parameter estimation in marked Hawkes processes, while \citet{Lambe2025} consider discretely observed Hawkes processes based on aggregated count data, and further explore a bootstrap bias correction procedure that requires no additional neural network training. 
In the spatial setting, \citet{Vihrs2022} employ convolutional neural networks (CNNs) combined with hand-crafted summary statistics to estimate parameters of spatial interaction models.
These three approaches are primarily aimed at producing point estimates of model parameters rather than posterior distributions.
More recent work has explored amortized full Bayesian inference using normalizing flows, through the availability of software packages such as BayesFlow \citep{Radev2022}. BayesFlow has been applied to LGCPs by \citet{wang2026inference} to enable fast generation of approximate posterior samples of model parameters. While the application of BayesFlow to LGCPs represents an important step toward full Bayesian inference for spatial point process models, the approach of \citet{wang2026inference}, like that of \citet{Vihrs2022}, relies on carefully constructed summary statistics of the observed point pattern data.

In the spatiotemporal setting, neural point-process models have been developed in which neural networks are used to flexibly parameterize the conditional intensity function or its components.
\citet{Chen2021} factorize the conditional spatiotemporal intensity into a temporal ground intensity and a conditional spatial density modeled using continuous normalizing flows.
In a related direction, \citet{Zhou2022} represent the spatiotemporal intensity through a latent kernel-based model whose parameters are learned using neural networks and whose integral is available in closed form.
\citet{Chukwuemeka2026} have recently developed a multivariate spatiotemporal neural Hawkes process that learns spatial and temporal decay dynamics, allowing flexible self- and cross-excitation and inhibition without prespecified triggering kernels.
These approaches incorporate neural networks into the point-process model itself for flexible event modeling and prediction.
In contrast, our interest lies in using neural networks to perform amortized Bayesian estimation of the interpretable parameters of a prespecified point-process model.

Existing amortized neural approaches to parameter estimation for point-process models exhibit several limitations.
Methods developed to date typically aggregate observations in spatial or temporal ``bins'', and they do not respect the continuous nature of observed spatiotemporal point patterns.
Those that do not aggregate instead rely on hand-crafted summary statistics or fixed-dimensional representations. While doing so allows an amortized network to be applied to datasets of varying size or extent through appropriate rescaling or normalization, these methods inevitably lead to information loss and to a reduction in statistical efficiency.
Moreover, training neural networks for point process data remains computationally demanding, as it requires repeated simulation of point pattern data over the entire domain of interest; this is particularly costly for large spatial domains or for models of dense point patterns.

In this work, we develop a methodology for amortized inference from point pattern data that circumvents the above limitations. Our method takes as input the raw point pattern realizations, accommodates both spatial and temporal dependence, and scales efficiently to large datasets.
The scaling capability is achieved through a divide-and-conquer neural Bayes estimation approach where large point pattern realizations are first partitioned into smaller, manageable blocks that are summarized locally, and that are then aggregated through a permutation-invariant attention mechanism.
By training the network on data generated over varying domain sizes, we enable it to generalize to substantially larger domains than those seen during training, with only a slight reduction in statistical efficiency.
This strategy can reduce computational and memory requirements while preserving fine-scale information, thereby enabling flexible and scalable inference for complex spatial and spatiotemporal point process models.
We also show that aggregating blockwise summaries, rather than blockwise parameter estimates, leads to improved estimators in a Bayes risk sense.


The remainder of the paper is organized as follows. Section~\ref{sec:NB} introduces neural Bayes estimation as the foundation of our approach. Section~\ref{sec:DnC_NB} develops the proposed divide-and-conquer neural Bayes estimator, which enables scalable inference for complex point process models through blockwise representation and aggregation. The section also provides theoretical support for aggregating blockwise summaries rather than blockwise parameter estimates. Section~\ref{sec:sim} evaluates our method through simulation studies for log-Gaussian Cox processes and spatiotemporal Hawkes processes.
Section~\ref{sec:real} demonstrates the practical utility of our proposed method through an application to whale call data, including parameter estimation and short-term prediction of whale calling activity.
Finally, Section~\ref{sec:discussion} concludes with a discussion of limitations and directions for future research.

\section{Neural Bayes Estimation}
\label{sec:NB}

The objective of point estimation is to recover the unknown parameter vector $\btheta \in \bTheta$ from observed data $\by \in \mathcal{Y}$ via an estimator $\hat{\btheta} : \mathcal{Y} \to \bTheta$, where $\mathcal{Y}$ denotes the sample space and $\bTheta$ denotes the parameter space. 
Let $L(\btheta,\hat{\btheta})$ be a nonnegative loss function, which is a function of the estimate $\hat{\btheta}$ and the true parameter $\btheta$. The Bayes risk of an estimator $\hat{\btheta}$ is defined as its prior-averaged risk,
\begin{align}
    \int_{\bTheta} \int_{\mathcal{Y}} 
    L\big(\btheta, \hat{\btheta}(\by)\big)
    f(\by \mid \btheta) p(\btheta) \, \text{d}\by \, \text{d}\btheta,
    \label{eq:bayes-risk}
\end{align}
where $p(\btheta)$ denotes a prior distribution on $\bTheta$, and 
$f(\cdot \mid \btheta)$ is the probability model of $\by$ given $\btheta$. Any estimator that minimizes \eqref{eq:bayes-risk} is referred to as a Bayes estimator under the loss $L(\cdot, \cdot)$ and prior $p(\cdot)$.
Bayes estimators are theoretically attractive due to their consistency and asymptotic efficiency under mild assumptions \citep[see][Theorems 5.2.4 and 6.8.3]{lehmann1998theory}, yet closed-form expressions for \eqref{eq:bayes-risk} are typically unavailable. 

Motivated by universal approximation theorems for neural networks \citep[e.g.,][]{hornik1989multilayer,zhou2020universality}, neural methods have been employed to approximate these estimators in complex models \citep{zammit2025neural}.
Let $\hat{\btheta}(\by;\bgamma)$ denote a neural network that produces  a point estimate from data $\by \in \mathcal{Y}$, where $\bgamma$  collects the network parameters. A neural approximation to the Bayes  estimator can be obtained by minimizing the empirical risk
\begin{align}
    \bgamma^* 
    = \argmin_{\bgamma}
    \frac{1}{K} \sum_{k=1}^{K} 
    L\!\left(\btheta^{(k)}, 
    \hat{\btheta}(\by^{(k)};\bgamma)\right),
    \label{eq:empirical-risk}
\end{align}
where $\btheta^{(k)} \sim p(\cdot)$ and  $\by^{(k)} \sim f(\cdot \mid \btheta^{(k)})$, independently for $k=1,\dots,K$.
The objective in \eqref{eq:empirical-risk} provides a Monte Carlo  approximation to the Bayes risk. The optimization problem can be solved efficiently via back-propagation and stochastic gradient-based methods. The trained network approximately minimizes the Bayes risk and is referred to as a \emph{neural Bayes estimator} \citep{Sainsbury-Dale2024,zammit2025neural}.

Neural Bayes estimators, like classical Bayes estimators, aim to approximate a particular point functional of the posterior distribution. The specific posterior summary that is targeted depends on the choice of loss function used during training. For example, minimizing a suitable surrogate for the 0–1 loss \citep[e.g.,][]{sainsbury2025neural} yields an estimator approximating the posterior mode, while absolute-error and squared-error losses correspond to the posterior median and mean, respectively. In addition, posterior quantiles can be obtained by adopting a quantile loss function, which enables the construction of credible intervals \citep{Sainsbury-Dale2025}.

Neural Bayes estimators offer several practical advantages over classical estimators. After solving the optimization problem in \eqref{eq:empirical-risk}, the trained network can be applied repeatedly to new datasets at negligible computational cost. This makes the method particularly suitable for settings in which the same model must be fitted repeatedly, as the initial training cost is effectively amortized through repeated uses. Also, neural Bayes estimators do not require evaluation of the likelihood function. Instead, they rely solely on the ability to simulate from the model, making them especially attractive for models with analytically or computationally intractable likelihoods. For such models, neural Bayes estimators can yield substantial improvements over classical approximate methods such as composite-likelihood methods \citep{Sainsbury-Dale2025}.


All likelihood-free inference methods discussed in Section~\ref{sec:intro}, such as ABC and neural inference approaches, rely fundamentally on repeated simulation of data. Such simulations can become computationally demanding when large spatial domains are considered or when data are densely observed. Moreover, training neural networks that operate directly on raw data may require substantial computational resources and memory, particularly in large-scale settings. 
For instance, one could consider using modern neural network architectures, such as transformers \citep{vaswani2017attention}, which are highly flexible models capable of capturing complex dependence structures by allowing each data point to interact with all others. However, applying such models directly to a point pattern would require computations over all pairs of events; this leads to a computational and memory complexity that is quadratic in the number of events. For large point patterns, this quickly becomes prohibitive.
To alleviate this computational burden, 
we incorporate a divide-and-conquer approach into the neural Bayes framework by partitioning the domain into blocks and aggregating the resulting blockwise summaries to estimate the global model parameters.
Our approach is similar to that of \citet{Hector2024} for geostatistical data, but differs in two important respects. First, rather than combining block-specific parameter estimates, we aggregate learned summary vectors that capture local point-pattern structure. Second, we use an additional neural network to optimize the global aggregation stage, rather than relying on a bootstrap-based weighted average.
In Section \ref{sec:sim_lgcp} we show how optimizing the aggregation in this way can lead to more statistically and computationally efficient estimators.
Details of our proposed approach are presented in the following section.

\section{Divide-and-Conquer Neural Bayes Estimation}
\label{sec:DnC_NB}

Let $\mathcal{D} \subset \mathbb{R}^p$ denote the observation domain, and let $\by$ denote an observed point pattern realization on $\mathcal{D}$. We consider $p=2$ for spatial point patterns and $p=3$ for spatiotemporal point patterns, where the latter corresponds to two spatial dimensions and one temporal dimension.
In the spatial setting, $\by \equiv \{\bs_i : i = 1, \dots, n\}$ denotes observed event locations in $\mathcal{D}$. In the spatiotemporal setting, the point pattern $\by \equiv \{(\bs_i, t_i) : i = 1, \dots, n \}$ consists of event times paired with spatial locations. For point pattern data, both the event locations (and times, when applicable) and the total number of events $n$ are random.

We consider settings in which the observed point pattern is dense (i.e., $n$ is large) and the domain $\mathcal{D}$ is large, for which inference on $\btheta$ becomes computationally demanding. In such cases, even simulation-based neural methods may incur substantial computational cost during training due to the need for repeated simulation of large point pattern data over $\mathcal{D}$. Training networks directly on raw data can further exacerbate computational and memory demands in large-scale settings.
In what follows, we present a divide-and-conquer neural Bayes estimation framework to address this challenge. 
Section~\ref{sec:block} introduces the domain partition, and Section~\ref{sec:divide} describes the construction of block-specific summary vectors.
Section~\ref{sec:conquer} details how we aggregate these block-specific summaries for global parameter estimation. 
Section~\ref{sec:varyDomain} describes how we account for varying domain configurations, and Section~\ref{sec:uncertainty} outlines our approach to uncertainty quantification using quantile estimation.
Finally, Section~\ref{sec:information} compares the information available from blockwise summaries and that from the corresponding blockwise parameter estimates, and gives a justification for aggregating summaries rather than estimates.
We present our proposed methodology in the spatiotemporal setting; the corresponding formulation for purely spatial point patterns follows as a straightforward special case by omitting the temporal component.

\subsection{Domain Partitioning}
\label{sec:block}

We partition the spatiotemporal domain $\mathcal{D} \subset \mathbb{R}^3$ into a collection of $m$ blocks, which may or may not overlap,
\begin{align}
    \mathcal{D} = \bigcup_{j=1}^m \mathcal{D}_j, \qquad \mathcal{D}_j \subset \mathcal{D}, \nonumber
\end{align}
where the block size and degree of overlap need to be specified.
Our method is generally insensitive to block size and overlap, except when the blocks are extremely small. The accuracy of the full-domain parameter estimates tend to improve slightly as the number of blocks increases, either through smaller blocks or greater overlap. However, when blocks are so small that they contain few or no points, estimation accuracy may deteriorate. The corresponding sensitivity analysis is reported in Appendix~\ref{app:lgcp_block_sensitivity}. We denote by $\by_j$ the subset of the point pattern that is in $\mathcal{D}_j$.
We introduce this domain partitioning purely for computational purposes, and do not impose any independence assumptions on the process across blocks. The partitioning allows local processing of the data, reduces memory requirements, and facilitates parallel computation.

\subsection{Blockwise Summary Representation}
\label{sec:divide}


Unlike \citet{Vihrs2022}, \citet{Lambe2025} and \citet{wang2026inference} who pre-specify summary statistics for use in neural estimation, here we use a neural network to extract informative summary statistics from each block, which we later combine in the conquering step.  We construct informative summary statistics by training a neural network to estimate process-model parameters from the block data. We let this neural network comprise a ``summary network'' and a linear module that transforms the summary statistics to parameter estimates. These estimates are discarded in our approach; they are solely used to inform what summary statistics are important for making inference on $\btheta$.

We simulate data to train the neural network as follows: For $k=1,\ldots,K_b$, we sample parameters from the prior distribution $\boldsymbol{\theta}^{(k)} \sim p(\cdot)$ and generate a point pattern $\by_1^{(k)} \equiv \{(\bs_i^{(k)}, t^{(k)}_i) : i = 1, \dots, n_k\}$ from the model $f(\cdot \mid \boldsymbol{\theta}^{(k)})$ on a \emph{single} spatiotemporal block $\mathcal{D}_1$. Here $\vert \mathcal{D}_1 \vert \ll \vert \mathcal{D} \vert$ and we choose the first block without loss of generality. The quantity $n_k$ denotes the number of events in the $k$th point pattern, and $K_b$ denotes the number of model simulations at the single block level.
We order the indices of the points in the point pattern according to their temporal coordinate; that is, $t^{(k)}_1 < t^{(k)}_2 < \cdots < t^{(k)}_{n_k}$.

We rescale the event coordinates of each point pattern to lie in $[-1,1]$ along each dimension. For a spatial or temporal coordinate $u \in [u_{\min},u_{\max}]$ (i.e., where $u$ represents either $s_1$, $s_2$, or $t$), we apply the transformation
\[
\tilde u = 2\,\frac{u - (u_{\min}+u_{\max})/2}{u_{\max}-u_{\min}},
\]
where $u_{\min}$ and $u_{\max}$ denote fixed lower and upper bounds of the coordinate $u$ over the cube $\mathcal{D}_1$.
Let $(\tilde{\bs}_i, \tilde{t}_i)$ denote the resulting rescaled coordinates.
Then $(\tilde{\bs}_i, \tilde{t}_i) \in \tilde{\mathcal{D}}_1 \equiv [-1,1]^3$.
We additionally include a temporal increment feature in the description of each event, $\log(1+\Delta \tilde{t}_i)$, where $\Delta \tilde{t}_i \equiv \tilde{t}_i - \tilde{t}_{i-1}$ and $\tilde{t}_0 = -1$. 
While the consideration of this feature is not a requirement, we found it to be helpful to infer parameters that capture the temporal dependence structure of the point pattern.
We collect the features of the events in
\begin{align}
    \bx_1^{(k)} \equiv \{ (\tilde{\bs}_i^{(k)}, \tilde{t}_i^{(k)}, \log(1+\Delta \tilde{t}_i^{(k)})): i = 1,\dots,n_k \},\quad k = 1,\dots,K_b, \label{eq:features}
\end{align}
which consists of the rescaled coordinates and transformed temporal increments. 


For the summary-extraction component of the network we use a transformer-based network \citep{vaswani2017attention}.
The transformer architecture is well suited for point pattern data because its self-attention mechanism enables it to capture interactions among events, such as clustering or self-excitation.
In standard transformer architectures, positional encodings are used to inject information about the ordering of the elements in a sequence. 
Here, we do not use positional encodings so that the output can be made permutation-invariant with respect to the ordering of events. 
This is appropriate when making inference on parameters for point patterns, which are (unordered) sets.\footnote{This permutation invariance with respect to the ordering of the events should not be confused with the time ordering used to construct $\Delta\tilde t_i^{(k)}$; the latter is a canonical operation needed only for feature construction.} 

The transformer architecture, which we denote by $h_{\bphi}(\cdot)$ and which has trainable parameters $\bphi$, maps the event features $\bx_1$ to a $d$-dimensional vector $\bz_1$ that captures important summaries of the point pattern on $\mathcal{D}_1$. The parameter estimate from data in the first block, $\hat{\btheta}_1 = \hat{\btheta}_b(\bx_1)$, is then obtained by applying a simple linear mapping $T_{\bdelta}(\cdot)$ with trainable parameters $\bdelta$ to the summary vector constructed from data in that block. Our neural Bayes estimator is therefore given by $\hat{\btheta}_b(\bx_1) = T_{\bdelta} ( \bz_1 )$ with $\bz_1 = h_{\bphi}(\bx_1)$.
Following the framework described in Section~\ref{sec:NB}, we train the network by minimizing the empirical risk
\begin{align}
(\bphi^*, \bdelta^*) 
= \argmin_{\bphi, \bdelta}
\frac{1}{K_b}\sum_{k=1}^{K_b} 
L_b\!\left(\btheta^{(k)},\, T_{\bdelta}(h_{\bphi}(\bx_1^{(k)}))\right), \label{eq:emprisk_summary}
\end{align}
where $L_b(\cdot,\cdot)$ denotes a suitable loss function such as the squared error loss. Details of the network architecture are provided in Appendix~\ref{sec:summaryNetwork}.

Our primary aim of training this neural network on data in a single block is not to estimate $\btheta$ but, rather, to extract the summary statistics from each block that are informative of $\btheta$. Crucially, in the aggregation stage we combine the (typically high-dimensional) informative summaries, and not the estimates themselves, to obtain a global estimate. Once trained, these informative summaries are obtained on each block practically instantaneously; this allows us to scale inference to large domains with relative ease.



\subsection{Global Aggregation of Blockwise Summaries}
\label{sec:conquer}

In this section, we introduce an aggregation function $g(\cdot)$ that maps the blockwise summary vectors to the ``global'' parameter estimate according to some loss function $L_g(\cdot, \cdot)$. Since the optimal aggregation is generally unknown, we approximate it using a neural network $g_{\bpsi}(\cdot)$, where $\bpsi$ denotes the collection of network parameters.
Assume, for now, that we are able to easily simulate point patterns on the entire domain $\mathcal{D}$ (this restriction will be relaxed in Section~\ref{sec:varyDomain}).
To train $g_{\bpsi}(\cdot)$, we generate training datasets as follows. For $k=1,\ldots,K_g$, where $K_g$ is the number of simulations, we draw parameters from the prior distribution $\btheta^{(k)} \sim p(\cdot)$ and simulate a point pattern $\by^{(k)} \sim f(\cdot \mid \btheta^{(k)})$ on 
the domain $\mathcal{D}$
using the same prior distribution and data-generating model as in Section~\ref{sec:divide}. 
We then partition each simulated dataset into $m$ blocks using the same block size as in the summary network training stage (Section~\ref{sec:divide}), yielding block-level data $\{\by_j^{(k)}\}_{j=1}^{m}$ with corresponding features $\{\bx_j^{(k)}\}_{j=1}^{m}$.
We then apply the trained summary network to each block to obtain summary vectors $\bz_j^{(k)} = h_{\bphi^*}(\bx_j^{(k)})$, which can be computed in parallel across blocks.

The relative position of the spatiotemporal region from which summaries are extracted is also informative of the parameter. Let $\bc_j$ denote the center coordinates of block $\mathcal{D}_j$. 
We use the extent of $\mathcal{D}$ to rescale these coordinates to $\tilde{\bc}_j \in [-1,1]^3$,
and append $\tilde{\bc}_j$ to each summary vector to form $\bZ^{(k)} = \{(\bz_j^{(k)}, \tilde{\bc}_j)\}_{j=1}^{m}$, which consists of informative summaries as well as the relative block position within $\mathcal{D}$.
We then use $\bZ^{(k)}$, $k = 1,\dots, K_g$, to train the aggregation network $g_{\bpsi}(\cdot)$, which we implement using a second
transformer-based architecture. 
The network defines the mapping $\hat{\btheta}_g(\bZ) = g_{\bpsi}\!\left(\bZ\right)$.
Following the framework described in Section~\ref{sec:NB}, we train $g_{\bpsi}(\cdot)$ by minimizing the empirical risk
\begin{align}
\bpsi^*
= \argmin_{\bpsi}
\frac{1}{K_g}\sum_{k=1}^{K_g}
L_g\!\left(
\btheta^{(k)},
g_{\bpsi}\!\left( \bZ^{(k)} \right)
\right),
\nonumber
\end{align}
where $L_g(\cdot,\cdot)$ denotes a suitable loss function. 

Akin to the summary stage, the transformer is suitable in this setting because it can capture interactions among blocks when forming the global representation. We do not use positional encodings, but instead we include the spatial information $\tilde{\bc}_j$ in the input features. This allows the network to learn spatiotemporal relationships among blocks directly from their relative position in the domain. 
We provide details of the network architecture used for $g_{\bpsi}(\cdot)$ in Appendix~\ref{sec:aggregator}.

\begin{figure}[!t]
\centering
\begin{tikzpicture}[
    x=1cm, y=1cm,
    >={Latex[length=3mm,width=2.2mm]},
    line/.style={line width=0.9pt},
    arrow/.style={->, line width=0.9pt},
    box/.style={draw, line width=0.9pt, minimum width=1.8cm, minimum height=0.95cm, align=center},
    smallbox/.style={draw, line width=0.9pt, minimum width=1.8cm, minimum height=0.95cm, align=center},
    bigbox/.style={draw, line width=0.9pt, minimum width=2cm, minimum height=6.5cm, align=center},
    every node/.style={font=\footnotesize}
]

\draw[line] (0,0) rectangle (4,4);
\draw[line] (2,0) -- (2,4);
\draw[line] (0,2) -- (4,2);

\foreach \x/\y in {
0.81/3.42, 1.13/3.02, 1.30/2.82, 1.00/2.62, 1.30/3.18, 1.50/2.96,
2.60/3.62, 3.36/3.42, 3.20/3.08, 3.12/2.74, 3.56/2.48,
2.26/2.74, 2.52/1.74, 2.70/1.62, 3.06/1.68, 3.55/1.86, 3.65/1.42, 1.80/1.84,
0.40/1.06, 0.80/0.74, 2.16/0.66
}{
  \fill (\x,\y) circle (2.3pt);
}

\node at (0.50,2.30) {$\by_1^{\text{obs}}$};
\node at (2.50,2.30) {$\by_2^{\text{obs}}$};
\node at (0.50,0.32) {$\by_3^{\text{obs}}$};
\node at (2.50,0.32) {$\by_4^{\text{obs}}$};

\node[box] (s1) at (5.5,4.53) {Summary\\Network};
\node[box, below=0.55cm of s1] (s2) {Summary\\Network};
\node[box, below=0.55cm of s2] (s3) {Summary\\Network};
\node[box, below=0.55cm of s3] (s4) {Summary\\Network};

\draw[arrow] (1.15,3.85) 
.. controls ($(s1.west)+(-1.5,1.2)$) and ($(s1.west)+(-0.5,0.5)$)
.. (s1.west);
\draw[arrow] (3.75,3.00) to[out=0, in=180] (s2.west);
\draw[arrow] (3.75,1.00) to[out=0, in=180] (s3.west);
\draw[arrow] (1.15,0.15) 
.. controls ($(s4.west)+(-1.5,-1.2)$) and ($(s4.west)+(-0.5,-0.5)$)
.. (s4.west);

\node[right=0.1cm of s1, yshift=0.25cm] (z1) {$\bz_1^{\text{obs}}$};
\node[right=0.1cm of s2, yshift=0.25cm] (z2) {$\bz_2^{\text{obs}}$};
\node[right=0.1cm of s3, yshift=0.25cm] (z3) {$\bz_3^{\text{obs}}$};
\node[right=0.1cm of s4, yshift=0.25cm] (z4) {$\bz_4^{\text{obs}}$};

\node[below=0.50cm of z1, xshift=-0.30cm] (c1) {$\tilde{\bc}_1$};
\node[below=0.50cm of z2, xshift=-0.30cm] (c2) {$\tilde{\bc}_2$};
\node[below=0.50cm of z3, xshift=-0.30cm] (c3) {$\tilde{\bc}_3$};
\node[below=0.50cm of z4, xshift=-0.30cm] (c4) {$\tilde{\bc}_4$};

\node[smallbox, right=1cm of s1] (l1) {Linear};
\node[smallbox, right=1cm of s2] (l2) {Linear};
\node[smallbox, right=1cm of s3] (l3) {Linear};
\node[smallbox, right=1cm of s4] (l4) {Linear};

\draw[arrow] (s1.east) -- (l1.west);
\draw[arrow] (s2.east) -- (l2.west);
\draw[arrow] (s3.east) -- (l3.west);
\draw[arrow] (s4.east) -- (l4.west);

\node[right=0.7cm of l1] (t1) {$\hat{\btheta}_1$};
\node[right=0.7cm of l2] (t2) {$\hat{\btheta}_2$};
\node[right=0.7cm of l3] (t3) {$\hat{\btheta}_3$};
\node[right=0.7cm of l4] (t4) {$\hat{\btheta}_4$};

\draw[arrow] (l1.east) -- (t1.west);
\draw[arrow] (l2.east) -- (t2.west);
\draw[arrow] (l3.east) -- (t3.west);
\draw[arrow] (l4.east) -- (t4.west);

\node[bigbox] (agg) at (11.6, 1.30) {Aggregation\\Network};

\draw[arrow] (z1.south) |- ($(agg.west |- c1.east)+(0,0.16cm)$);
\draw[arrow] (z2.south) |- ($(agg.west |- c2.east)+(0,0.16cm)$);
\draw[arrow] (z3.south) |- ($(agg.west |- c3.east)+(0,0.16cm)$);
\draw[arrow] (z4.south) |- ($(agg.west |- c4.east)+(0,0.16cm)$);

	\draw[arrow] (c1.east) -- (c1.east -| agg.west);
	\draw[arrow] (c2.east) -- (c2.east -| agg.west);
	\draw[arrow] (c3.east) -- (c3.east -| agg.west);
	\draw[arrow] (c4.east) -- (c4.east -| agg.west);

\node[right=0.7cm of agg.east] (global) {$\hat{\btheta}$};
\draw[arrow] (agg.east) -- (global.west);

\end{tikzpicture}
\caption{Schematic illustration of our proposed divide-and-conquer neural Bayes estimation framework. 
For simplicity, the domain is illustrated in two dimensions.}
\label{fig:sketch}
\end{figure}
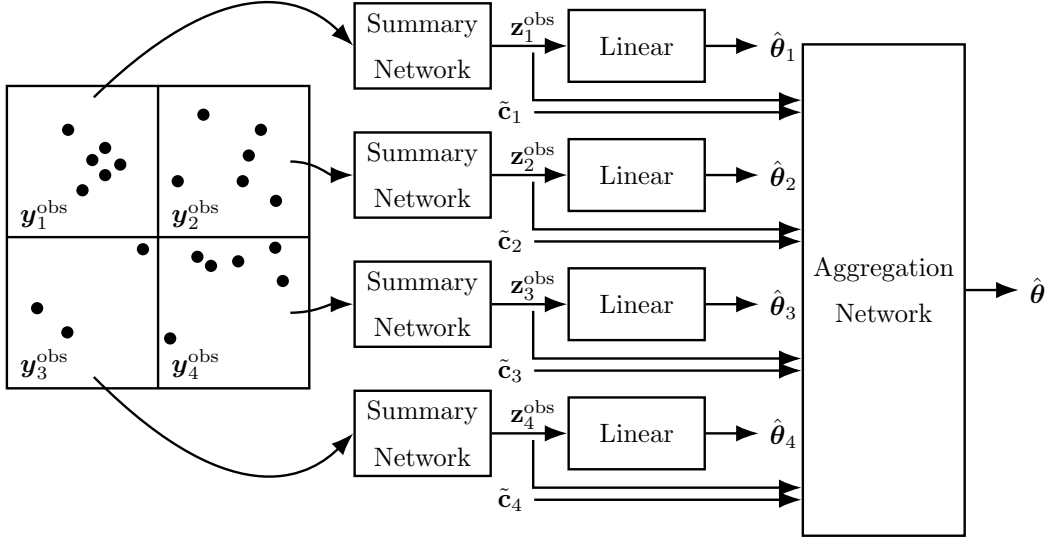

Figure~\ref{fig:sketch} provides a schematic overview of our proposed divide-and-conquer neural Bayes estimation framework, here shown for a two-dimensional domain partitioned into four blocks for simplicity.
Consider an observed point pattern 
$\by^{\text{obs}}$ on $\mathcal{D}$
and let $\{\by_j^{\text{obs}}\}_{j=1}^{m}$ and $\{\bx_j^{\text{obs}}\}_{j=1}^{m}$ denote the resulting blockwise point patterns and the corresponding features, respectively. The summary network $h_{\bphi^*}(\cdot)$ is applied to each block to obtain summary vectors $\bz_j^{\text{obs}} = h_{\bphi^*}(\bx_j^{\text{obs}}), j = 1,\dots,m$.
The spatial block reference $\tilde{\bc}_j$ is appended to each summary vector to form the inputs $ \bZ^{\text{obs}} = \{(\bz_j^{\text{obs}}, \tilde{\bc}_j)\}_{j=1}^{m}$.  These inputs are then fed into the aggregation network $g_{\bpsi^*}(\cdot)$ to produce the final parameter estimate $\hat{\btheta} = g_{\bpsi^*}\!\left( \bZ^{\text{obs}} \right)$.


This aggregation step combines locally learned blockwise summaries to recover the global model parameters. By operating on compact block-level representations rather than the full point pattern, the approach substantially reduces computational complexity. 
This hierarchical representation enables scalable inference for large spatial or spatiotemporal point pattern datasets while preserving dependence information across the domain.
Once trained, the summary and aggregation networks can be applied repeatedly to new datasets without retraining, enabling efficient amortized inference.

\subsection{Variable Domain Size and Configuration}
\label{sec:varyDomain}

For our neural Bayes estimator to be approximately Bayes over a range of domain sizes and configurations, we treat the domain configuration as a random object $\mathcal C$ taking values in a set $\mathfrak C$ of all admissible domain configurations. We then augment the Bayes risk associated with the aggregation network to
\begin{align}
\sum_{c \in \mathfrak C} P(\mathcal C=c)
\int_{\bTheta}\int_{\mathcal Y_{c}}
L_g\!\left(
\btheta,\;
g_{\bpsi}\!\left(
\left\{
\big(h_{\bphi^*}(\bx_j), \tilde{\bc}_j\big)
\right\}_{j=1}^{m(c)}
\right)
\right)
f(\by_{c}\mid \btheta)\,p(\btheta)\, \text{d} \by_{c}\, \text{d} \btheta,
\label{eq:BayesRisk_varyingDomainConfig}
\end{align}
where $\bx_j$ is constructed from $\by_c$ as shown in \eqref{eq:features}, and where each configuration $c \in \mathfrak C$ specifies both the number and the spatiotemporal arrangement of blocks. For a given configuration $c$, $m(c)$ denotes the corresponding number of blocks, $\mathcal Y_{c}$ denotes the space of all possible point patterns defined on the domain induced by $c$, and $\by_{c} = \cup_{j=1}^{m(c)} \by_j$ denotes the collection of point patterns across these blocks.

In this case, an additional step is required in the training stage of the aggregation network. 
Let $\mathcal{D}_{\text{max}} \equiv \mathcal{D}_S \times \mathcal{D}_T$ denote the largest domain we can feasibly simulate point pattern data on, which can be smaller or larger than $\mathcal{D}$.
For $k = 1,\ldots,K_g$, we then construct a domain configuration $\mathcal{C}$ by first sampling the number of temporal blocks uniformly from $\{1, \dots, M_T\}$, where $M_T$ denotes the maximum allowable number of blocks in the temporal domain $\mathcal{D}_T$, and then sampling the number of spatial blocks uniformly from $\{1, \dots, M_S\}$, where $M_S$ denotes the maximum allowable number of blocks in the spatial domain $\mathcal{D}_S$. For simplicity, we arrange spatial blocks to maximize the number of shared boundaries, resulting in configurations that are approximately rectangular.

Training the aggregation network under varying configurations yields estimators that generalize well across different sizes and arrangements. Importantly, it also allows the aggregation network to learn the effect of domain size on the optimal way to combine the summaries, allowing it to extrapolate effectively to domains larger than those encountered during training. This is particularly useful 
when generating training data on the entire domain is computationally infeasible. Simulation results 
showing that extrapolation to larger domains is indeed possible with only a slight loss in statistical efficiency
are presented in Section~\ref{sec:sim_varyDomain}.

\subsection{Uncertainty Quantification}
\label{sec:uncertainty}

We also use the aggregation network introduced in Section~\ref{sec:conquer} to estimate posterior quantiles. 
For a scalar parameter $\theta$, 
an estimate of the posterior $\tau$-quantile, $q^{(\tau)}$, can be found by minimizing
the quantile (pinball) loss
\begin{align*}
L_{\tau}(\theta, q^{(\tau)}) 
= 
(q^{(\tau)} - \theta) \left[ \mathbb{I}(q^{(\tau)} > \theta) - \tau \right],
\end{align*}
where $\tau \in (0,1)$ is the desired posterior quantile level, and $\mathbb{I}(\cdot)$ denotes the indicator function. 
For a parameter vector $\btheta = (\theta_1, \dots, \theta_{d_{\btheta}})^\top$ and quantile level $\tau$, we consider the joint loss $L_{Q_\tau}(\btheta, \boldsymbol{q}^{(\tau)}) = \sum_{i=1}^{d_{\btheta}} L_{\tau}(\theta_i, q^{(\tau)}_i)$, under which the Bayes estimator is given by the vector of marginal posterior $\tau$-quantiles \citep[Theorem~2]{Sainsbury-Dale2025}.

To estimate multiple quantile levels $( \tau_1, \dots, \tau_R)$ while ensuring monotonicity (i.e., $\tau_1<\tau_2<\cdots<\tau_R$), we adopt a modified version of the construction in \citet{Sainsbury-Dale2025}. 
We express the aggregation network for the $r$th quantile using an attention unit $\nu_{\boldsymbol{\xi}}(\cdot)$ with parameters $\boldsymbol{\xi}$, and $r$ multilayer perceptrons (MLPs) with parameters $\boldsymbol{\zeta}_1,\dots,\boldsymbol{\zeta}_r$. We construct the quantile estimators sequentially as
\begin{align}
\boldsymbol{q}^{(\tau_1)}(\bZ) 
&= 
\mathrm{MLP}^{(\tau_1)}_{\bzeta_1}\!\left( \nu_{\boldsymbol{\xi}}(\bZ) \right), \nonumber\\
\boldsymbol{q}^{(\tau_r)}(\bZ) 
&= 
\boldsymbol{q}^{(\tau_1)}(\bZ)
+
\sum_{j=2}^r 
\sigma\!\left( 
\mathrm{MLP}^{(\tau_j)}_{\bzeta_j}\!\left( \nu_{\boldsymbol{\xi}}(\bZ) \right)
\right),
\quad r = 2,\dots,R,
\nonumber
\end{align}
where $\sigma(\cdot)$ is a nonnegative activation function not bounded from above that is applied elementwise (e.g., softplus). We collect the parameters for the combined quantile (aggregation) network in $\boldsymbol{\psi} = (\boldsymbol{\zeta}_1^\top,\dots,\boldsymbol{\zeta}_R^\top,\boldsymbol{\xi}^\top)^\top$.
This construction guarantees monotonicity of the estimated quantiles across levels. All quantile-specific networks share the same representation $\nu_{\boldsymbol{\xi}}(\cdot)$, while separate MLPs are used to capture quantile-specific features. Compared to training independent networks for each quantile level, this approach reduces computational cost by sharing information across quantiles.
The model parameters are learned by minimizing the additive loss
\begin{align}
L_Q\!\left(\btheta, \boldsymbol{q}^{(\tau_1)}(\bZ), \dots, \boldsymbol{q}^{(\tau_R)}(\bZ) \right)
=
\sum_{r=1}^R \sum_{i=1}^{d_{\boldsymbol{\theta}}}
L_{\tau_r}\!\left(
\theta_i,\;
{q}^{(\tau_r)}_i(\bZ)
\right),
\label{eq:loss_quantile}
\end{align}
where ${q}^{(\tau_r)}_i(\bZ)$ denotes the $i$th component of the $\tau_r$th quantile estimator. 

\subsection{Bayes-Risk Comparison for Blockwise Aggregation}
\label{sec:information}

We now compare, in terms of Bayes risk, the difference between aggregating blockwise summaries and aggregating the corresponding blockwise parameter estimates. Throughout this section, we treat the trained maps $h_{\bphi^*}(\cdot)$ and $T_{\bdelta^*}(\cdot)$ as fixed.

For a domain configuration $\mathcal C$, define the collections of blockwise summaries and blockwise parameter estimates, respectively, as
\begin{align*}
\bZ_{\mathcal C}
&=
\left\{
\big(h_{\bphi^*}(\bx_j),\tilde{\bc}_j\big)
\right\}_{j=1}^{m(\mathcal C)},
\\
\bU_{\mathcal C}
&=
\left\{
\big(T_{\bdelta^*}\{h_{\bphi^*}(\bx_j)\},\tilde{\bc}_j\big)
\right\}_{j=1}^{m(\mathcal C)}.
\end{align*}
The following proposition shows that the attainable Bayes risk when using $\bZ_{\mathcal C}$ will always be equal to, or lower than, the attainable Bayes risk when using $\bU_{\mathcal C}$.

\begin{proposition}
\label{prop:general_information}
Let $L(\btheta,\hat{\btheta})$ be a nonnegative measurable loss function, where $\hat{\btheta}\in\bTheta$ denotes a candidate estimate. Define
\begin{align*}
R_Z^*(L)
&=
\inf_{d_Z(\cdot,\cdot)}
\mathbb E\left[
L\!\left(
\btheta,d_Z(\bZ_{\mathcal C},\mathcal C)
\right)
\right],
\\
R_U^*(L)
&=
\inf_{d_U(\cdot,\cdot)}
\mathbb E\left[
L\!\left(
\btheta,d_U(\bU_{\mathcal C},\mathcal C)
\right)
\right],
\end{align*}
where $d_Z(\cdot,\cdot)$ and $d_U(\cdot,\cdot)$ respectively map $(\bZ_{\mathcal C},\mathcal C)$ and $(\bU_{\mathcal C},\mathcal C)$ to estimates of $\btheta$, where the infima are taken over all measurable estimators, and where the expectations are taken under the joint distribution of $(\btheta,\by_{\mathcal C}, \mathcal C)$. Then
\[
R_Z^*(L)\leq R_U^*(L).
\]
\end{proposition}

A proof of Proposition~\ref{prop:general_information} is given in Appendix~\ref{sup:sec:bayes_risk_monotonicity}. 
The proof relies on the fact that any estimator based on $\bU_{\mathcal C}$ can be recovered from $\bZ_{\mathcal C}$, and that the converse is not true in general, since the blockwise estimates may discard information in the summaries that is relevant to $\btheta$. The result also holds for dependent or overlapping blocks. The following corollary characterizes the difference between the two optimal Bayes risks under the ubiquitous case of squared-error loss.

\begin{corollary}
\label{cor:squared_information}
Suppose that $\mathbb E\|\btheta\|^2<\infty$, and define
\[
\mathbf{m}_Z
=
\mathbb E(\btheta\mid\bZ_{\mathcal C},\mathcal C),
\qquad
\mathbf{m}_U
=
\mathbb E(\btheta\mid\bU_{\mathcal C},\mathcal C).
\]
Then the Bayes squared-error risks satisfy
\begin{align*}
R_U^*-R_Z^*
=
\mathbb E\left[
\|\mathbf{m}_Z-\mathbf{m}_U\|^2
\right]
\geq 0.
\end{align*}
Equality holds if and only if $\mathbf{m}_Z=\mathbf{m}_U$ almost surely.
\end{corollary}

A proof of Corollary~\ref{cor:squared_information} is given in Appendix~\ref{sup:sec:squared_information_gap}. The proof relies on the fact that the sigma-algebra generated by $\{\bU_{\mathcal C}, \mathcal{C}\}$ is a subset of that generated by $\{\bZ_{\mathcal C}, \mathcal{C}\}$. 

Proposition~\ref{prop:general_information} compares the Bayes risks over all measurable estimators. In practice, optimization over an unrestricted class is infeasible. We instead minimize the empirical risk over flexible neural-network classes. The following remark relates the risks of the fitted estimators to the corresponding Bayes risks.

\begin{remark}
\label{rem:learned_risk}
For a loss $L(\cdot,\cdot)$, let $\hat{\btheta}_Z=g_{\bpsi^*}(\bZ_{\mathcal C})$ and $\hat{\btheta}_U=a^*(\bU_{\mathcal C})$ denote fitted blockwise-summary-based and blockwise-estimate-based rules, respectively. Let $R(\hat{\btheta};L)=\mathbb E\{L(\btheta,\hat{\btheta})\}$. The excess risks of the fitted rules based on their respective inputs are
\begin{align*}
\mathcal E_Z(L)
&=
R(\hat{\btheta}_Z;L)-R_Z^*(L),\\
\mathcal E_U(L)
&=
R(\hat{\btheta}_U;L)-R_U^*(L).
\end{align*}
The risk difference then satisfies
\begin{align*}
R(\hat{\btheta}_U;L)-R(\hat{\btheta}_Z;L)
&=
\{R_U^*(L)-R_Z^*(L)\}
+\{\mathcal E_U(L)-\mathcal E_Z(L)\}.
\end{align*}
\end{remark}
A derivation of this risk decomposition is given in Appendix~\ref{sup:sec:fitted_rules}.
The first term, $R_U^*(L)-R_Z^*(L)$, is the reduction in the Bayes risk achieved by retaining the blockwise summaries. The fitted summary-based rule has a lower risk exactly when this reduction exceeds $\mathcal E_Z(L)-\mathcal E_U(L)$, the difference between the two excess risks.
Therefore, the reduction in Bayes risk obtained by using blockwise summaries instead of blockwise estimates may be offset by differences in the excess risks of the fitted rules. However, this is unlikely to be the case when flexible neural network architectures are used in the aggregation stage, and our experiments in Section~\ref{sec:sim_lgcp} provide empirical evidence for a gain in statistical efficiency when aggregating summaries instead of estimates.

\par

\section{Simulation Experiments}
\label{sec:sim}

In this section, we conduct several simulation studies to evaluate the performance of our proposed divide-and-conquer neural Bayes estimator for spatial and spatiotemporal point processes. In Section~\ref{sec:sim_lgcp}, we estimate the parameters of a spatial log-Gaussian Cox process model for spatial point pattern realizations. We compare 
inferences on $\btheta$ to those from Markov chain Monte Carlo (MCMC), 
which is regarded as the gold standard.
We also compare our method to
other neural inference approaches, including BayesFlow as implemented in \citet{wang2026inference}, the method of \citet{Vihrs2022}, and the black-box divide-and-conquer approach of \citet{Hector2024}. 
In Section~\ref{sec:sim_hawkes}, we consider a spatiotemporal Hawkes process model for spatiotemporal point pattern data. 
We only compare our inferences on $\btheta$ to those from MCMC since,
to our knowledge, no neural Bayesian inference approaches have been developed for spatiotemporal point pattern data.
In Section~\ref{sec:sim_varyDomain}, we demonstrate the importance of accounting for variability in domain configuration during training to achieve both flexibility and scalability. 

Implementation details of the neural networks are provided in Appendix~\ref{app:implementation}.
For training the summary network, we simulate 100,000 training datasets and 2,000 validation datasets. For the aggregation network, we simulate 100,000 training datasets, 2,000 validation datasets, and 400 test datasets. Data generation is performed in parallel. To mitigate edge effects, we simulate point patterns on an extended domain and we only retain events within the domain of interest.
All experiments are conducted on the Duke Compute Cluster, using a node equipped with an Intel Xeon Gold 6252 CPU and an NVIDIA RTX 6000 Ada Generation GPU. 
Code for reproducing the results shown in this manuscript will be available at \url{https://github.com/bokgyeong/DnCNBE} (code available with submission for review). 

\subsection{Spatial Log-Gaussian Cox Process}
\label{sec:sim_lgcp}

In this section, we consider a spatial log-Gaussian Cox process (LGCP) model whose stochastic intensity at a spatial location $\bs$ in a bounded region $\mathcal{D}_S \subset \mathbb{R}^2$ is defined as
\begin{align}
\lambda_{L}(\bs) &= \exp \left\{ \gamma + w(\bs) \right\}, \nonumber
\end{align}
where $w(\bs)$ is a Gaussian process (GP) with mean zero and exponential covariance function
$C(\bs,\bs'; \kappa,\rho) = \kappa e^{ -\Vert \bs - \bs' \Vert / \rho }$
for $\bs, \bs' \in \mathcal{D}_S$. Here, $\kappa > 0$ denotes the variance and $\rho > 0$ is a spatial range parameter of the latent GP. 
Given an observed point pattern $\by = \{\bs_1, \dots, \bs_n \}$, the likelihood function can be written as
\begin{align*}
    \mathcal{L}(\lambda_{L}(\cdot) \mid \by) \propto \exp\left\{ - \int_{\mathcal{D}_S} \lambda_{L}(\bs) \text{d}\bs \right\} \prod_{i = 1}^n \lambda_{L}(\bs_i).
\end{align*}
This likelihood function is computationally intractable since it marginalizes over the latent process $w(\cdot)$.

\begin{figure}[!tb]
    \centering
    \includegraphics[width=0.8\linewidth]{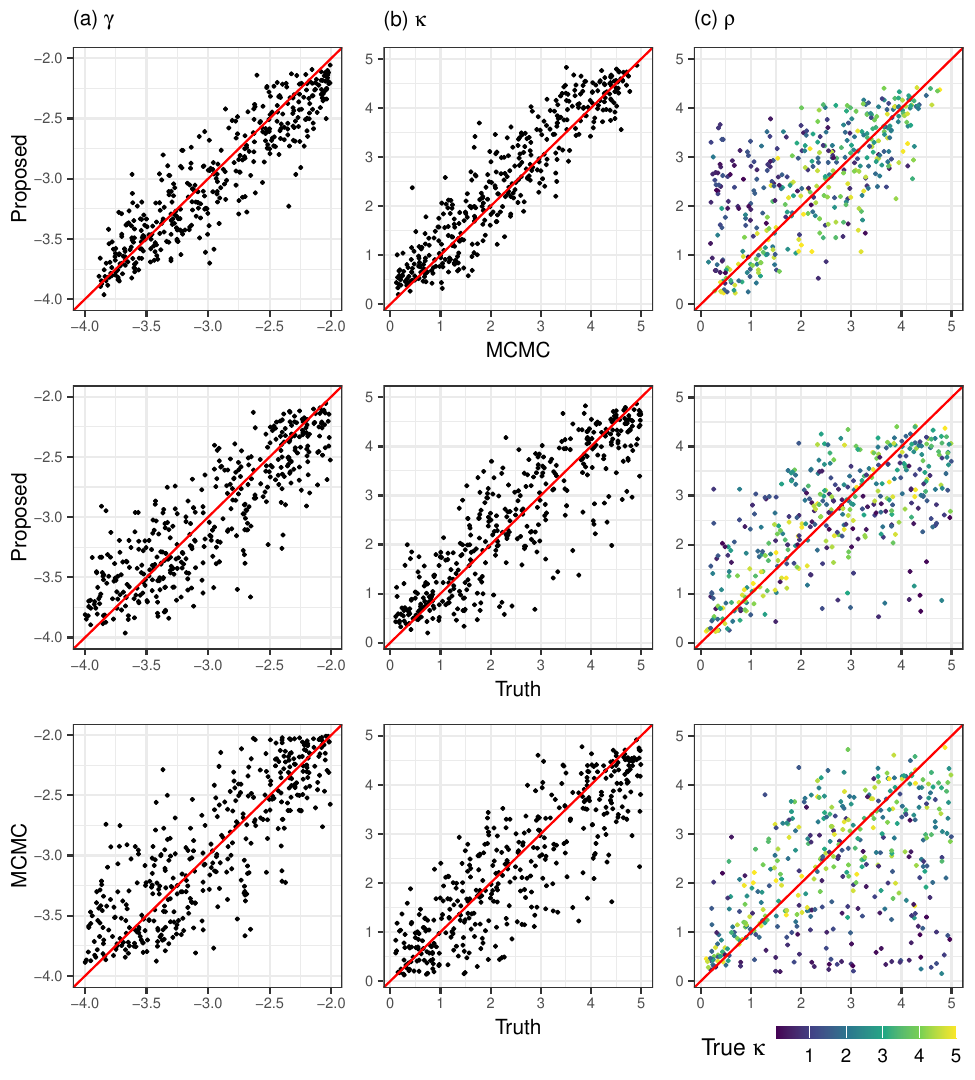}
    \caption{Comparison of posterior median estimates from our proposed estimator and MCMC with each other and with the true parameter values, based on 400 test datasets for the LGCP model. Columns correspond to parameters. (Top) Proposed versus MCMC. (Middle) Proposed versus truth. (Bottom) MCMC versus truth. For $\rho$, the color scale indicates the values of $\kappa$.}
    \label{fig:lgcp_proposed_mcmc_truth}
\end{figure}

We let $\gamma \sim \text{Unif}(-4, -2)$, $\kappa \sim \text{Unif}(0.1, 5)$, and $\rho \sim \text{Unif}(0.1, 5)$.  
We partition the spatial domain into square blocks of side length $10$ with an overlap of $3$ between adjacent blocks.
We consider the domain $\mathcal{D}_S = [0, 24] \times [0, 24]$, yielding a total of $9$ blocks.

We simulate datasets on $\mathcal{D}_1 = [0, 10] \times [0, 10]$ to train the summary network, with the number of observed points ranging from 0 to 2,448. 
On average, generating a single dataset takes approximately 1.21 seconds. 
For the aggregation network, we simulate datasets under varying domain configurations, with $\mathcal{D}_S$ representing the largest domain, rather than over a single fixed domain.
This results in sample sizes ranging from 0 to 6,580, and an average simulation time of 13.54 seconds per dataset,
which is manageable in our computing environment (in Section~\ref{sec:sim_varyDomain}, we show that the estimator can extrapolate to larger domains than those seen during training if needed).
The total training time for our proposed estimator is 5.2 minutes, comprising 4.6 minutes for the summary network and 37 seconds for the aggregation network. In our implementation, MCMC requires approximately 109 hours to run 30,000 iterations for a single dataset (largely due to the repeated factorization of the Gaussian process covariance matrix),
whereas our trained networks output parameter estimates in approximately 0.002 seconds. 

\setlength{\tabcolsep}{3.5pt} 
\begin{table}[t]
\caption{Comparison of estimation performance for our proposed method, the approach of \citet{Vihrs2022}, BayesFlow \citep{wang2026inference}, and the divide-and-conquer method of \citet{Hector2024}. Reported metrics include the root mean squared error (RMSE) between point estimates and true parameter values, empirical coverage of nominal 95\% intervals, and their average lengths. Training and inference times are also reported.}
\label{tab:lgcp_compar}
\begin{center}
\begin{tabular}{llrrrrr}
\toprule
 Parameter & Method & RMSE & Coverage & \makecell{Avg.\\Length} & \makecell{Training\\(sec)} & \makecell{Inference\\(sec)} \\
\midrule
   & Proposed & 0.29 & 0.955 & 1.10 & 313 &   0.002 \\ 
  $\gamma$ & Vihrs (2022) & 0.30 & 0.970 & 1.11 & 51 &   0.002 \\  
   & BayesFlow & 0.33 & 0.953 & 1.29 & 259 & 0.142 \\ 
   & Hector and Lenzi (2026) & 0.39 & 0.172 & 0.26 & 276 & 0.026 \\ 
   \midrule
   & Proposed & 0.70 & 0.948 & 2.64 &  \\ 
  $\kappa$ & Vihrs (2022) & 0.69 & 0.943 & 2.52 &   \\ 
   & BayesFlow & 0.74 & 0.955 & 3.14 & \\ 
   & Hector and Lenzi (2026) & 0.98 & 0.238 & 0.77 & \\ 
   \midrule
   & Proposed &  1.04 & 0.948 & 3.38 &  \\ 
  $\rho$ & Vihrs (2022) & 0.99 & 0.948 & 3.32 &   \\ 
   & BayesFlow & 1.04 & 0.953 & 3.60 & \\ 
   & Hector and Lenzi (2026) & 1.23 & 0.135 & 0.50 & \\ 
\bottomrule
\end{tabular}    
\end{center}
\end{table}

We jointly estimate the 2.5\%, 50\%, and 97.5\% marginal posterior quantiles using the loss \eqref{eq:loss_quantile}.
Figure~\ref{fig:lgcp_proposed_mcmc_truth} compares posterior median estimates from our proposed estimator and MCMC with the true parameter values across 400 test datasets generated on the full nine-block domain $\mathcal{D}_S$. For $\gamma$ and $\kappa$, both methods recover the true parameter values well. 
Compared to the other parameters, the range parameter $\rho$ exhibits substantially greater variability. This is consistent with the well-known difficulty of estimating correlation range parameters in spatial point process settings \citep{Diggle2013}. Discrepancies between the estimates and the true values are particularly pronounced when $\kappa$ is close to zero, with larger deviations observed for the MCMC estimates. This behavior is expected, as little information about $\rho$ is available when the variance parameter $\kappa$ of the Gaussian process is near zero, corresponding to a negligible effect of the latent process.

We also compare our method with existing neural network–based inference approaches for spatial point processes, including those of \citet{Vihrs2022} and \citet{wang2026inference}. Both approaches rely on pre-defined summary statistics rather than raw point pattern inputs. In this example, we use the same set of $55$ summary statistics as in \citet{wang2026inference} for both methods. The approach of \citet{Vihrs2022} employs a convolutional neural network (CNN) to obtain point estimates of the parameters, whereas \citet{wang2026inference} uses normalizing flows to obtain the posterior samples of the parameters.
We further compare our method with the divide-and-conquer framework of \citet{Hector2024}, originally developed for geostatistical data. To adapt their approach to spatial point pattern data, we use the trained network \(T_{\hat{\bdelta}}(h_{\hat{\bphi}}(\bx_j))\) in \eqref{eq:emprisk_summary} to obtain block-specific parameter estimates $\hat{\btheta}_j$, $j=1,\ldots,m$. We then combine these estimates using the bootstrap-based weighted aggregation procedure proposed by \citet{Hector2024}. Following their procedure, we first compute the overall mean estimate $\hat{\btheta}_m=m^{-1}\sum_{j=1}^m\hat{\btheta}_j$. For each bootstrap replicate, data on every block are generated from the model parameterized by $\hat{\btheta}_m$, and the resulting blockwise bootstrap estimates are used to construct the aggregation weights. We generate $B = 2,000$ bootstrap replicates in parallel across blocks.
Table~\ref{tab:lgcp_compar} shows that our method, the approach of \citet{Vihrs2022}, and \citet{wang2026inference}'s BayesFlow perform similarly in terms of the root mean squared error (RMSE) between posterior median estimates and the true parameter values, as well as empirical coverage of 95\% credible intervals. The proposed method and that of \citet{Vihrs2022} yield comparable average credible-interval lengths, whereas BayesFlow produces slightly wider credible intervals.
In contrast, the divide-and-conquer approach of \citet{Hector2024}, which yields a frequentist point estimate and confidence interval, exhibits larger RMSE values, substantially lower empirical coverage, and shorter interval lengths across all parameters. These results highlight the potential benefits of retaining informative blockwise summaries rather than reducing them to blockwise parameter estimates, and are consistent with Proposition~\ref{prop:general_information} and Remark~\ref{rem:learned_risk}. 

For models such as the LGCP, where informative low-dimensional summary statistics are available, our method performs comparably to the approach of \citet{Vihrs2022}, while the latter is computationally more efficient during training. However, such informative summaries are not available for all point process models. In particular, for spatiotemporal Hawkes processes, no widely accepted set of low-dimensional summary statistics exists.
In contrast, our method does not rely on the availability of such summaries and remains applicable across a broader class of models. We illustrate this advantage in the next section through a spatiotemporal Hawkes process example.

\subsection{Spatiotemporal Hawkes Process}
\label{sec:sim_hawkes}

\begin{figure}[!t]
    \centering
    \includegraphics[width=0.99\linewidth]{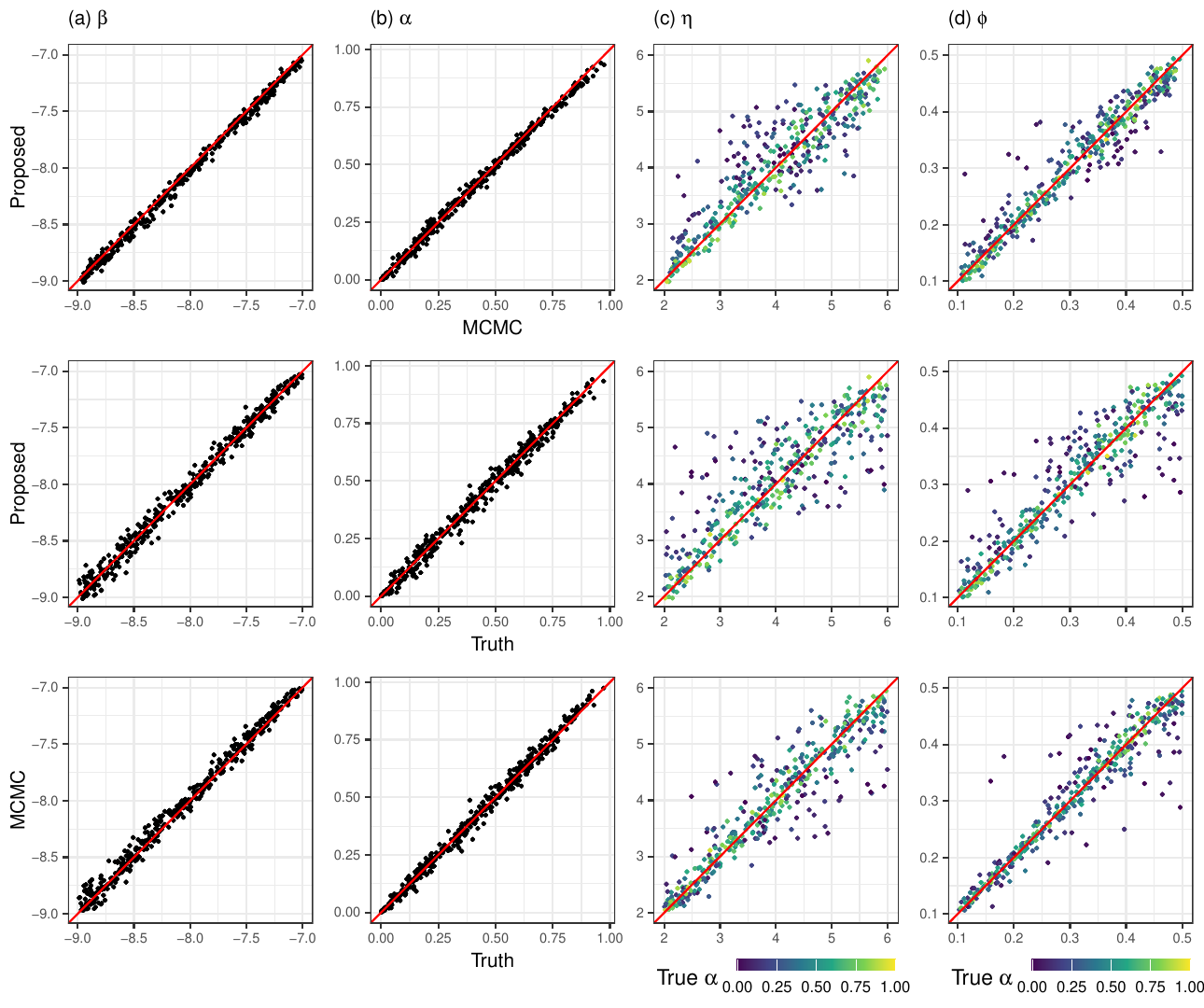}
    \caption{Comparison of posterior median estimates from our proposed estimator and MCMC with each other and with the true parameter values, based on 400 test datasets for the spatiotemporal Hawkes process model. Columns correspond to parameters. (Top) Proposed posterior versus MCMC. (Middle) Proposed versus truth. (Bottom) MCMC versus truth. For $\eta$ and $\phi$, the color scale indicates the values of $\alpha$.}
    \label{fig:sthawkes_proposed_mcmc_truth}
\end{figure}

Consider now a spatiotemporal point pattern $\by = \{ (\bs_i, t_i) : i = 1, \dots, n \}$, where $\bs_i \in \mathcal{D}_S \subset \mathbb{R}^2$ denotes the location of the $i$th event and $t_i \in \mathcal{D}_T \subset \mathbb{R}$ denotes its occurrence time. Let $\bh_t = \{ (\bs_i, t_i) : t_i < t \}$ denote the history of the process up to time $t$. 
The conditional intensity of a spatiotemporal Hawkes process at location $\bs$ and time $t$ is given by
\begin{align}
\lambda_H(\bs, t \mid \bh_t)
=
\exp(\beta)
+
\frac{\alpha}{2 \pi \eta \phi^2}
\sum_{i: t_i < t}
\exp\!\left(-\frac{t-t_i}{\eta}\right)
\exp\!\left(-\frac{\|\bs-\bs_i\|}{\phi}\right),
\nonumber
\end{align}
where $\exp(\beta)$ denotes the background intensity corresponding to events that occur independently of the past history, and the second term represents the excitation intensity that captures the contribution of past events to the current intensity.
The temporal decay parameter $\eta > 0$ controls the rate at which the influence of past events diminishes over time, while the spatial decay parameter $\phi > 0$ governs the spatial range of interaction. The parameter $\alpha > 0$ controls the magnitude of the excitation effect. For the exponential kernel considered here, $\alpha$ coincides with the branching ratio, defined as the expected number of offspring events generated by a single event. The condition $\alpha < 1$ ensures that the process remains stable, in the sense that the total number of events does not diverge over time.

We let $\beta \sim \mathrm{Unif}(-9, -7)$, $\alpha \sim \mathrm{Unif}(0, 1)$, $\eta \sim \mathrm{Unif}(2, 6)$, and $\phi \sim \mathrm{Unif}(0.1, 0.5)$. 
We partition $\mathcal{D}_S$ into square blocks of side length $10$ with a spatial overlap of $1$.
Similarly, we partition $\mathcal{D}_T$ into segments of length $600$ with a temporal overlap of $5$.
We consider the domain $\mathcal{D} = \mathcal{D}_S \times \mathcal{D}_T = [0, 28] \times [0, 37] \times (0, 1195]$, whose complete partition contains $24$ blocks.


For training the summary network, we simulate datasets on $\mathcal{D}_1 = [0, 10] \times [0, 10] \times (0, 600]$, with the number of observed points ranging from 1 to 2,444 and an average generation time of 6.8 seconds per dataset. 
For the aggregation network, we simulate datasets under varying domain configurations, with $\mathcal{D}$ representing the largest domain, rather than over a single fixed domain. This yields sample sizes between 1 and 7,565 and an average simulation time of 24.5 minutes per dataset.
The total training time for our proposed estimator is 19.7 minutes, comprising 18.1 minutes for the summary network and 1.6 minutes for the aggregation network. In our implementation, MCMC requires, on average, approximately 64 minutes to run 30,000 iterations for a single dataset, whereas our proposed estimator produces parameter estimates in approximately 0.002 seconds. After training, inference for new datasets can be performed with negligible computational cost.

\begin{table}[!tb]
\caption{Comparison of posterior estimation performance for our method and MCMC across 400 test datasets. Reported metrics include the root mean squared error (RMSE) between posterior median estimates and true parameter values, empirical coverage of 95\% credible intervals, and their average lengths.
}
\label{tab:sthawkes_compar}
\begin{center}
\begin{tabular}{llrrr}
\toprule
 Parameter & Method & RMSE & Coverage & \makecell{Avg. Length} \\\midrule
  \multirow{2}{*}{$\beta$}& Proposed & 0.06 & 0.94 & 0.25 \\
  & MCMC & 0.06 & 0.89 & 0.20 \\\midrule
   \multirow{2}{*}{$\alpha$} & Proposed & 0.03 & 0.95 & 0.11 \\
  & MCMC & 0.02 & 0.96 & 0.09 \\\midrule
   \multirow{2}{*}{$\eta$} & Proposed & 0.58 & 0.94 & 1.93\\
  & MCMC & 0.42 & 0.94 & 1.35 \\\midrule
   \multirow{2}{*}{$\phi$} & Proposed & 0.04 & 0.94 & 0.13\\
  & MCMC & 0.03 & 0.93 & 0.08 \\
\bottomrule
\end{tabular}
\end{center}
\end{table}


We jointly estimate the 2.5\%, 50\%, and 97.5\% marginal posterior quantiles using the loss \eqref{eq:loss_quantile}. Figure~\ref{fig:sthawkes_proposed_mcmc_truth} compares posterior median estimates from our method and MCMC with the true parameter values across 400 test datasets generated under the same varying domain configurations.
For $\beta$ and $\alpha$, both approaches closely recover the true values, and our proposed estimates align well with those from MCMC.
For the decay parameters $\eta$ and $\phi$, the estimates generally align well with the identity line, indicating good recovery of the true values. Some increased dispersion is observed when $\alpha$ is close to zero, which is expected, as the excitation component becomes negligible in this regime and provides limited information about $\eta$ and $\phi$.
Table~\ref{tab:sthawkes_compar} compares our method with MCMC in terms of the root mean squared error (RMSE) of posterior median estimates, together with the empirical coverage and average length of 95\% credible intervals. The two methods exhibit comparable performance in RMSE, although our method shows a slightly larger RMSE for $\eta$. Our approach tends to produce slightly wider credible intervals on average, while achieving empirical coverage close to the nominal 95\% level across all parameters. 

Overall, our proposed method achieves estimation accuracy comparable to MCMC, while providing reliable uncertainty quantification. 
Although our method requires an upfront computational cost for training, it enables amortized inference, yielding near-instantaneous predictions (approximately 0.002 seconds per dataset), in contrast to MCMC, which must be rerun from scratch for each new dataset (requiring several days for the LGCP model and approximately one hour for the spatiotemporal Hawkes model).

\subsection{Training with Varying Domain Configurations}
\label{sec:sim_varyDomain}

\begin{figure}[!t]
    \centering
    \includegraphics[width=0.85\linewidth]{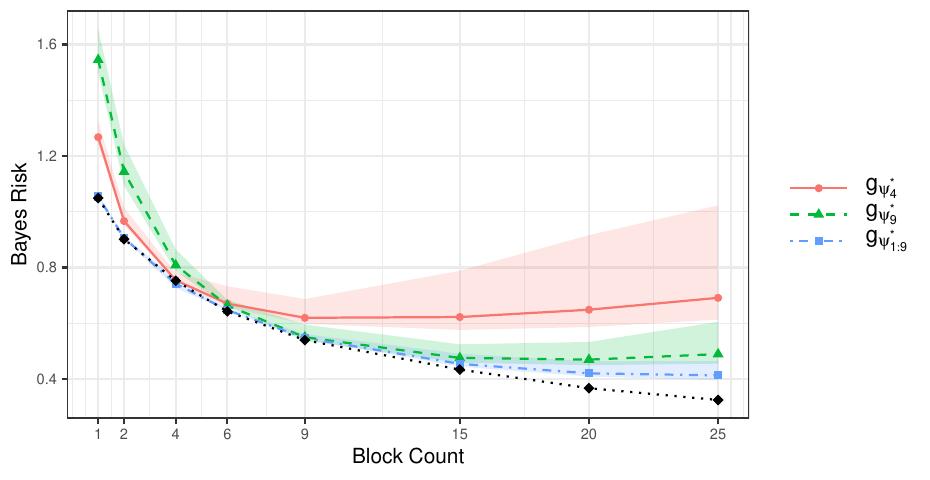}
    \caption{
    Bayes risk as a function of the number of blocks for estimators trained under different domain configurations. Solid, dashed, and dash-dotted lines correspond to estimators trained on domains with $4$, $9$, and $1$--$9$ blocks ($g_{\bpsi^*_4}$, $g_{\bpsi^*_9}$, and $g_{\bpsi^*_{1:9}}$), respectively. The black dotted line approximates the configuration-specific Bayes optimal risk. Points denote averages over 30 independent training runs, and shaded regions indicate the corresponding variability.
    }
    \label{fig:sim_varyDomain}
\end{figure}

Our proposed estimator based on the Bayes risk \eqref{eq:BayesRisk_varyingDomainConfig} can be applied to datasets with arbitrary domain sizes and configurations. However, its performance depends on the domain configuration used during training. To investigate this dependence and assess the benefits of training under varying domain sizes and configurations, we consider the LGCP example in Section~\ref{sec:sim_lgcp} and train estimators under different domain configurations. 
Specifically, $g_{\bpsi^*_4}(\cdot)$ and $g_{\bpsi^*_9}(\cdot)$ are aggregation networks trained on datasets generated over domains with exactly $4$ and $9$ blocks, respectively, while $g_{\bpsi^*_{1:9}}(\cdot)$ is trained on datasets generated over domains with varying configurations, with the number of blocks ranging from $1$ to $9$.

We generate test datasets under domain configurations with selected block counts between $1$ and $25$ (with 25 blocks corresponding to a much larger domain than that seen during training).
For each configuration, we evaluate the Bayes risk of the estimators $g_{\bpsi^*_4}(\cdot)$, $g_{\bpsi^*_9}(\cdot)$, and $g_{\bpsi^*_{1:9}}(\cdot)$, and approximate the ``optimal'' Bayes risk by training an aggregation network under that specific configuration. Figure~\ref{fig:sim_varyDomain} shows that the estimators $g_{\bpsi^*_4}(\cdot)$ and $g_{\bpsi^*_9}(\cdot)$ attain this approximate optimum when the test datasets have $4$ and $9$ blocks, respectively, matching the domain configurations used during training. For other block counts, their risks are higher than the approximate optimal risks, indicating a loss of efficiency.
By contrast, $g_{\bpsi^*_{1:9}}(\cdot)$ attains nearly optimal risk throughout the range of block counts represented during training, namely from $1$ to $9$. For larger block counts, its risk exceeds the configuration-specific approximate optimum, as expected, but the deterioration is modest. This indicates that the estimator generalizes reasonably well to larger domain configurations beyond those encountered during training.
See Section S2 of \citet{Sainsbury-Dale2024} for an analogous simulation experiment exploring the benefits of considering variable sample size during training.

\section{Application to Whale Call Data}
\label{sec:real}

\begin{figure}[!t]
    \centering
    \includegraphics[width=0.55\linewidth]{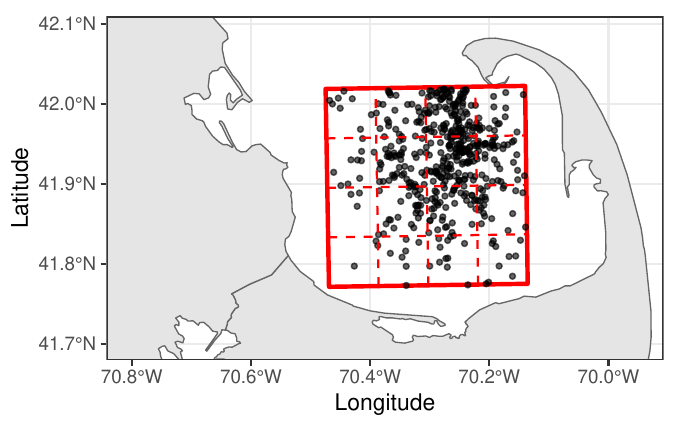}
    \caption{Whale call locations in Cape Cod Bay, MA, aggregated over the period from 2009-02-18 08:00:00 to 2009-02-19 18:00:00 UTC. The red box indicates the spatial region used for model fitting, and the dashed lines show the domain partition used in the divide-and-conquer framework. For visualization purposes, the blocks are displayed without overlap, although overlapping blocks are used in practice.
    }
    \label{fig:ccb}
\end{figure}



We apply our approach to analyze upcall data from North Atlantic right whales (\textit{Eubalaena glacialis}, hereafter NARW) collected in Cape Cod Bay, Massachusetts, USA, in 2009 \citep{palmerAccountingLombardEffect2022}. Upcalls are frequency-modulated signals used for communication among individuals \citep{clarkAcousticRepertoireSouthern1982}. Hydrophones are commonly deployed to document the presence of NARW throughout their range \citep{davisLongtermPassiveAcoustic2017}.
\citet{Kang2025whaleHawkes} analyze similar data consisting of the times at which individual NARW upcalls are received at each of 10 hydrophones using a multivariate Hawkes process model. In that framework, the background intensity describes contact calls, interpreted as initial calls aimed at establishing contact, while the excitation intensity captures potential countercalls, corresponding to responses triggered by previous upcalls.

With an array of multiple hydrophones, the arrival times of calls at different hydrophones can be used to infer the location of the calling whale \citep{watkinsSoundSourceLocation1972a}. We work with the estimated spatial locations of individual upcalls. 
We perform detection and localization using PAMGuard software \citep{gillespiePAMGuardApplicationSoftware2026}, which automatically matches detections across hydrophones that are inferred to come from the same upcall, and estimates the location of the sound source. Only upcalls that were loud enough to be detected on at least four hydrophones are retained for analysis.
The analysis is restricted to the period from 2009-02-18 08:00:00 to 2009-02-19 18:00:00 UTC, during which whale calls were observed frequently, resulting in 484 recorded calls. Figure~\ref{fig:ccb} shows the call locations aggregated over this period, with the red box indicating the spatial region used for model fitting.

We model the data using the spatiotemporal Hawkes process described in Section~\ref{sec:sim_hawkes}. We let $\beta \sim \mathrm{Unif}(-11.5, -8.2)$, $\alpha \sim \mathrm{Unif}(0, 1)$, $\eta \sim \mathrm{Unif}(0.1, 40)$, and $\phi \sim \mathrm{Unif}(0.1, 3)$. 
Under the prior for $\beta$, the expected number of background events ranges from 18.9 to 513.6.
We choose the prior ranges for $\eta$ and $\phi$ to be relatively broad compared to those used in the simulation study. 
We simulate training datasets from the model under the specified priors. We partition $\mathcal{D}_S = [0, 28] \times [0, 28]$ km into square blocks with side length $10$ km and spatial overlap $4$ km, while we partition $\mathcal{D}_T = (0, 2040]$ min into segments of length $540$ min with temporal overlap $40$ min, yielding a total of $64$ blocks.
In this case, we are able to simulate point patterns over the entire domain $\mathcal{D} = \mathcal{D}_{\text{max}}$ for training the aggregation network.



\begin{table}[!tb]
\caption{Posterior quantile summaries for the spatiotemporal Hawkes model based on our proposed method and MCMC.
\label{tab:real_parameter}}
\begin{center}
\begin{tabular}{llrrr}
\toprule
Parameter & Method & 2.5\% & Median & 97.5\% \\
\midrule
  $\beta$  & Proposed & -10.59 & -10.01 & -9.58 \\ 
   & MCMC & -10.28 & -9.89 & -9.57 \\  \midrule
  $\alpha$ & Proposed & 0.72 & 0.84 & 0.95 \\ 
   & MCMC & 0.80 & 0.90 & 0.98 \\  \midrule
  $\eta$ & Proposed & 11.35 & 14.95 & 20.39 \\ 
   & MCMC & 14.92 & 19.57 & 26.11 \\  \midrule
  $\phi$ & Proposed & 1.01 & 1.27 & 1.59 \\ 
   & MCMC & 1.17 & 1.36 & 1.58 \\ 
\bottomrule
\end{tabular}
\end{center}
\end{table}

Beyond parameter estimation, an important question is whether the fitted model is able to capture the short-term spatiotemporal dynamics of whale calling activity. We therefore evaluate the predictive performance of the fitted Hawkes process model through one-step-ahead intensity prediction and predictive scoring. We also compare its predictive performance to that of a spatial log-Gaussian Cox process model fitted at the previous time point. This serves as a strawman ``baseline'' that should be outperformed by any forecasting model.

Once training is complete, we compute the 2.5\%, 50\%, and 97.5\% marginal posterior quantile estimates for the model parameters from the observed data. 
For comparison, we also fit the model using MCMC with 20,000 iterations.
Table~\ref{tab:real_parameter} shows that our proposed method yields estimates comparable to those obtained from MCMC.
The posterior median estimate of the excitation magnitude parameter $\alpha$ is $0.84$, with a $95\%$ credible interval of $(0.72, 0.95)$. This implies strong evidence of countercalling, as the interval is far away from zero. 
The posterior median estimate of the temporal decay parameter $\eta$ is $14.95$, with a $95\%$ credible interval of $(11.35, 20.39)$. 
This corresponds to an expected response time of approximately 15 minutes for NARW upcalls. This estimate is broadly consistent with the average inter-call interval of 17.5 minutes reported by \citet{davis2025upcalling} for NARWs in Southern New England.
For the spatial decay parameter $\phi$, the posterior median is $1.27$, with a $95\%$ credible interval of $(1.01, 1.59)$. 
This suggests that the spatial extent of responses (the distance at which the excitation effect decays to about 5\% of its initial magnitude) is approximately $3\phi \approx 3.8$ km. 

\begin{figure}[!tb]
    \centering
    \includegraphics[width=0.95\linewidth]{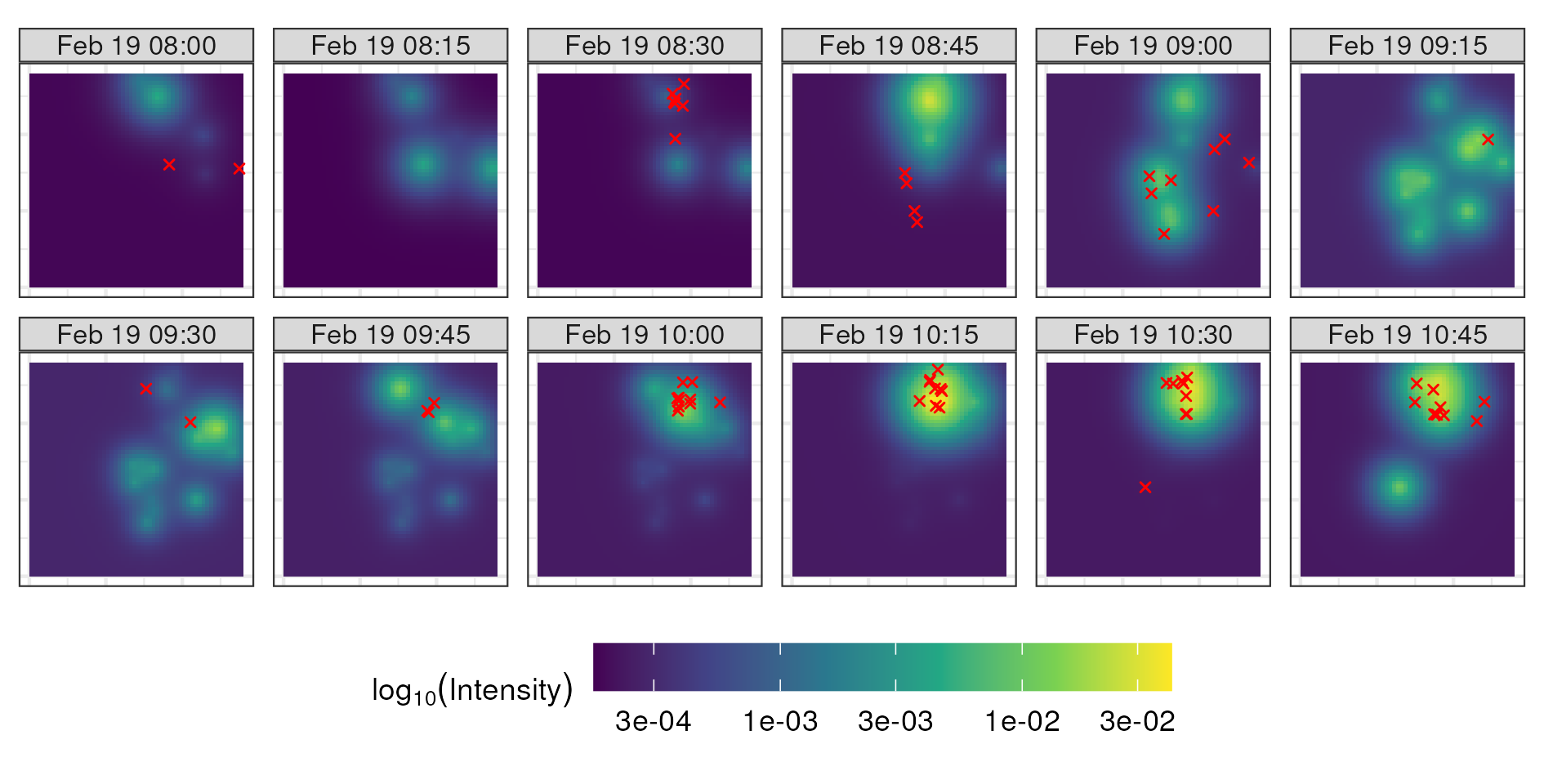}
    \caption{Selected 15-minute-ahead predicted intensity surfaces and corresponding observed data (red crosses) over consecutive intervals on 2009-02-19 for whale call data.}
    \label{fig:whale_pred}
\end{figure}

We now assess the ability of the spatiotemporal Hawkes process model to capture short-term spatiotemporal variation in whale calling activity.
We partition the study period $\mathcal{D}_T$ into $I$ equally sized subintervals,
\[
\mathcal{D}_{T,i} = (i\Delta t, (i+1)\Delta t], \qquad i = 0, \dots, I-1,
\]
where $I\Delta t = \max(\{t_i\}_i)$. 
For each interval $\mathcal{D}_{T,i}$, we predict the spatial intensity surface conditional on the observed event history up to time $i\Delta t$, denoted by $\bh_{i\Delta t}$, and the posterior median estimates of the model parameters $\hat{\btheta}_{i} = (\hat{\beta}_i, \hat{\alpha}_i, \hat{\eta}_i, \hat{\phi}_i)$. These parameter estimates are obtained using our proposed estimator fitted to the observations contained in the collection of overlapping temporal blocks whose most recent block ends at $i\Delta t$. Since inference with the trained estimator is essentially instantaneous, the intensity surface can be updated rapidly as new observations become available.
A forecast of the intensity at spatial location $\bs$ over $\mathcal{D}_{T,i}$ is given by
\begin{align}
&\int_{\mathcal{D}_{T,i}}
\lambda_H(\bs,u \mid \bh_{i\Delta t}; \hat{\btheta}_i)
\, \mathrm{d}u 
=
\Delta t e^{\hat{\beta}_i}
\nonumber  +\frac{\hat{\alpha}_i}{2\pi \hat{\phi}_i^2}
\sum_{j : t_j < i\Delta t}
e^{
-\frac{\|\bs-\bs_j\|}{\hat{\phi}_i}
}
\left[
e^{
-\frac{i\Delta t-t_j}{\hat{\eta}_i}
}
-
e^{
-\frac{(i+1)\Delta t-t_j}{\hat{\eta}_i}
}
\right].
\nonumber
\end{align}
Figure~\ref{fig:whale_pred} displays selected predicted intensity surfaces over consecutive future intervals of length $\Delta t = 15$ minutes on 2009-02-19. The complete set of predictions over the full study period is provided in Appendix~\ref{app:whale-intensity}.
Under the Hawkes process model, recent events exert a stronger influence on the current intensity surface through the temporal excitation kernel, so the predicted intensity surface is primarily determined by recent calling activity prior to $i\Delta t$. Overall, the predicted high-intensity regions align closely with the subsequently observed whale call locations, suggesting that the fitted model captures the short-term spatiotemporal dynamics of whale calling behavior reasonably well.

\begin{figure}
    \centering
    \includegraphics[width=0.75\linewidth]{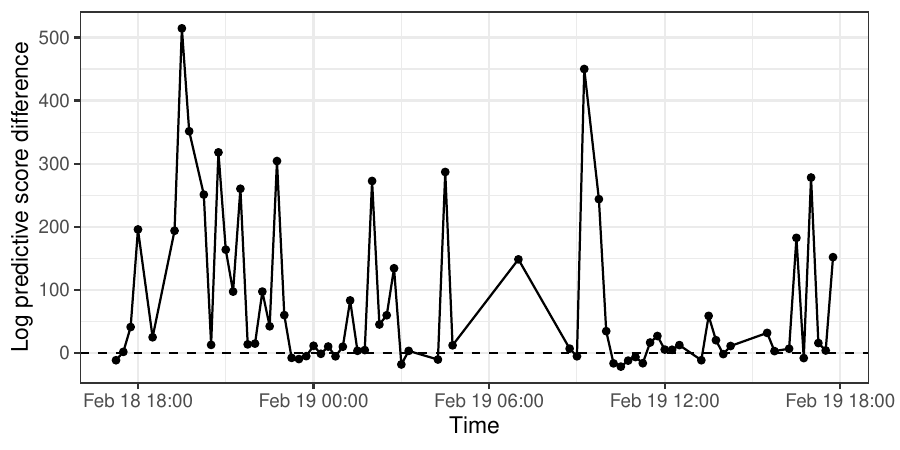}
    \caption{Difference in log predictive score between the Hawkes process and the LGCP (Hawkes $-$ LGCP) over consecutive 15-minute intervals.}
    \label{fig:whale_logratio}
\end{figure}

We further compare the predictive performance of the spatiotemporal Hawkes process model and the ``persistent forecast'' from the spatial LGCP model using one-step-ahead log predictive scores. For the Hawkes process model, we compute an inhomogeneous Poisson approximation to the predictive density over interval $\mathcal{D}_{T,i}$, for $i = 1, \dots, I-1$, which is given by
\begin{align*}
&p_H\!\left(
\{(\bs_j,t_j)\}_{j:t_j \in \mathcal{D}_{T,i}}
\mid
\bh_{i\Delta t};
\hat{\btheta}_i
\right)\\
&\propto
\exp\left\{
-
\int_{\mathcal{D}_S}
\int_{\mathcal{D}_{T,i}}
\lambda_H(
\bs,u
\mid
\bh_{i\Delta t};
\hat{\btheta}_i
)
\mathrm{d}u \mathrm{d}\bs
\right\}
\prod_{j:t_j \in \mathcal{D}_{T,i}}
\lambda_H(
\bs_j,t_j
\mid
\bh_{i\Delta t};
\hat{\btheta}_i
).
\end{align*}
For the LGCP model, we fit the model using observations from the preceding interval $\mathcal{D}_{T,i-1}$ and obtain the corresponding 
posterior mean estimate of the intensity surface, denoted by $\hat{\lambda}_{L, i-1}(\cdot)$, using the \texttt{lgcp} package in \texttt{R} \citep{taylor2015bayesian}.
To ensure stable estimation, the model is fitted and included in the comparison only when at least two observations are available over $\mathcal{D}_{T,i-1}$.
The resulting estimated intensity surface is then treated as the predicted surface over $\mathcal{D}_{T,i}$. The predictive density over interval $\mathcal{D}_{T,i}$ is given by
\begin{align*}
&p_L\!\left(
\{\bs_j\}_{j:t_j \in \mathcal{D}_{T,i}};
\hat{\lambda}_{L, i-1}(\cdot)
\right)
\propto
\exp\left\{
-
\int_{\mathcal{D}_S}
\hat{\lambda}_{L, i-1}(\bs)
\,\mathrm{d}\bs
\right\}
\prod_{j:t_j \in \mathcal{D}_{T,i}}
\hat{\lambda}_{L, i-1}(\bs_j).
\end{align*}
Figure~\ref{fig:whale_logratio} displays the difference in log predictive score between the Hawkes process and the LGCP. Positive values indicate stronger predictive support for the Hawkes model relative to the LGCP model. The difference is positive for most time intervals, often substantially so, suggesting that 
the temporal dynamics present in the spatiotemporal Hawkes process model are important for predicting whale calls.

This study demonstrates the practical advantages of amortized inference for model checking and comparison. For the spatiotemporal Hawkes process considered here, the parameter estimates required to compute the log predictive scores were obtained in under one second using our divide-and-conquer approach, whereas the equivalent MCMC analysis would have required more than 10 hours.

\section{Discussion}
\label{sec:discussion}

We have developed a divide-and-conquer neural Bayes estimation framework for scalable inference from large, heterogeneous point pattern data. 
Our approach partitions the domain into blocks that may or may not overlap, extracts informative summaries within each block using a neural network, and aggregates these summaries through a second neural network to estimate the model parameters. This construction reduces the computational and memory burden associated with large datasets, while retaining sufficient information for accurate inference.
Once trained, the resulting estimator can be applied to new datasets with negligible computational cost. This stands in contrast to likelihood-based approaches that must be executed separately for each dataset. 
Our empirical results indicate that our proposed method provides inference comparable to MCMC across a range of settings. They further suggest that training under varying domain configurations leads to estimators that remain robust across different domain sizes and arrangements, including larger domains than those observed during training, with only limited degradation in performance.

Our proposed framework is well suited to applications requiring repeated or large-scale fitting of spatial or spatiotemporal point process models, particularly when near real-time inference is required. Such settings arise in ecology (e.g., animal occurrence and interaction processes), epidemiology (e.g., modeling the spread of infectious diseases), seismology (e.g., earthquake occurrence modeling), neuroscience (e.g., analysis of neural spike train data), and criminology (e.g., modeling the occurrence of crime events). In these contexts, the ability to perform fast, amortized inference while accommodating complex dependence structures makes our proposed approach particularly appealing.

Several directions for future research arise from our proposed framework. First, while our proposed framework is developed for point pattern data, the underlying divide-and-conquer neural Bayes formulation is more broadly applicable and could be extended to other types of spatiotemporal data, such as geostatistical (point-referenced) data or spatiotemporal areal data. Exploring such extensions would further expand the scope of our method. 
Second, incorporating covariate information remains challenging, as our proposed approach relies on simulated training data. In particular, both model parameters and covariates must be specified or generated in order to simulate data from the model. While this is straightforward when covariates can be generated as part of the model specification, it becomes more difficult for externally observed covariates that cannot be readily generated under the model, and would require additional methodological development.
Finally, the current implementation focuses on estimating marginal posterior summaries rather than approximating the full posterior distribution. Extending the framework to learn the full posterior, for example by combining divide-and-conquer strategies with flexible density estimators such as normalizing flows \citep[e.g.,][]{Radev2022}, is a promising direction for future research.

\section*{Acknowledgements}
We thank Christopher Clark from Cornell University and Tina Yack from Duke University for collecting, processing and sharing the NARW data.
BK was supported by the National Research Foundation of Korea (NRF) grant funded by the Korea government (MSIT) (RS-2026-25487765). 
RSS and EMS were supported in part by awards (N000142312562, N000142412501) from the US Office of Naval Research.



\appendix

\section{Architecture of the Summary Network}
\label{sec:summaryNetwork}

Let $\bx_1 = \{(\tilde{\bs}_i,\tilde{t}_i,\log(1+\Delta\tilde{t}_i))^\top\}_{i=1}^{n}\in\mathbb{R}^{4}$ denote the sequence of event feature vectors of a spatiotemporal point pattern, where $n$ is the number of points.
We first map each input vector $(\tilde{\bs}_i,\tilde{t}_i,\log(1+\Delta\tilde{t}_i))^\top$ to a $d$-dimensional embedding, $\be_i=\mathbf{W}_e(\tilde{\bs}_i,\tilde{t}_i,\log(1+\Delta\tilde{t}_i))^\top+\bb_e$, where $\mathbf{W}_e\in\mathbb{R}^{d\times4}$ and $\bb_e\in\mathbb{R}^d$ are learnable parameters.
We also prepend a learnable summary vector $\be_0\in\mathbb{R}^d$ to the sequence of embeddings, forming
\begin{align*}
\bE=
\begin{pmatrix}
\be_0^\top\\
\be_1^\top\\
\vdots\\
\be_N^\top
\end{pmatrix}
\in\mathbb{R}^{(N+1)\times d},
\end{align*}
where $N$ is the maximum number of points that the neural network will be able to work with (here taken to be the maximum over all the simulated datasets used for training the network).
If $n<N$, the remaining embeddings, $\be_{n+1},\ldots,\be_N$, are zero-padded, and a binary mask is used to exclude them from subsequent computations.

We process the embedded sequence using a transformer encoder \citep{vaswani2017attention} consisting of $L$ layers.
Let $\bH^{(0)}=\bE$. For $\ell=1,\ldots,L$, the $\ell$th layer computes
\begin{align}
\tilde{\bH}^{(\ell)}
&=
\mathrm{LayerNorm}\!\left(
\bH^{(\ell-1)}
+
\mathrm{MHA}\!\left(\bH^{(\ell-1)}\right)
\right),
\label{eq:transformer}\\
\bH^{(\ell)}
&=
\mathrm{LayerNorm}\!\left(
\tilde{\bH}^{(\ell)}
+
\mathrm{MLP}\!\left(\tilde{\bH}^{(\ell)}\right)
\right).
\nonumber
\end{align}
Here, $\mathrm{MHA}(\cdot)$ denotes multi-head self-attention and $\mathrm{MLP}(\cdot)$ denotes a multi-layer perceptron applied independently to each input representation (i.e., each row of its argument).
The operator $\mathrm{LayerNorm}(\cdot)$ denotes layer normalization \citep{ba2016layer}, which normalizes each input representation across its feature dimension.

We now describe in detail the operations performed within the transformer encoder in \eqref{eq:transformer}.
Let $n_h$ denote the number of attention heads and $d_k$ the feature dimension of each attention head.
For layer $\ell$ and attention head $h$, let
\[
\mathbf{W}_{Q}^{(\ell,h)},
\mathbf{W}_{K}^{(\ell,h)},
\mathbf{W}_{V}^{(\ell,h)}
\in\mathbb{R}^{d\times d_k}
\]
denote learnable projection matrices. The query, key, and value matrices are
\[
\mathbf{Q}^{(\ell)}_h
=
\mathbf{H}^{(\ell-1)}\mathbf{W}_Q^{(\ell,h)},
\qquad
\mathbf{K}^{(\ell)}_h
=
\mathbf{H}^{(\ell-1)}\mathbf{W}_K^{(\ell,h)},
\qquad
\mathbf{V}^{(\ell)}_h
=
\mathbf{H}^{(\ell-1)}\mathbf{W}_V^{(\ell,h)},
\]
where $\mathbf{Q}^{(\ell)}_h,\mathbf{K}^{(\ell)}_h,\mathbf{V}^{(\ell)}_h\in\mathbb{R}^{(N+1)\times d_k}$.
The output of the $h$th attention head is
\begin{align}
\mathrm{head}^{(\ell)}_h
&=
\mathrm{softmax}\!\left(
\frac{\mathbf{Q}^{(\ell)}_h\mathbf{K}_h^{(\ell)\top}}{\sqrt{d_k}}
\right)
\mathbf{V}^{(\ell)}_h
\in\mathbb{R}^{(N+1)\times d_k}.
\nonumber
\end{align}
The outputs from all attention heads are concatenated along the feature dimension,
\[
\mathrm{Concat}\!\left(
\mathrm{head}^{(\ell)}_1,\ldots,\mathrm{head}^{(\ell)}_{n_h}
\right)
\in\mathbb{R}^{(N+1)\times n_hd_k},
\]
and projected back to the original feature dimension,
\begin{align}
\mathrm{MHA}\!\left(\mathbf{H}^{(\ell-1)}\right)
&=
\mathrm{Concat}\!\left(
\mathrm{head}^{(\ell)}_1,\ldots,\mathrm{head}^{(\ell)}_{n_h}
\right)
\mathbf{W}^{(\ell)}_O
\in\mathbb{R}^{(N+1)\times d},
\nonumber
\end{align}
where $\mathbf{W}^{(\ell)}_O\in\mathbb{R}^{n_hd_k\times d}$ denotes a learnable output projection matrix.
The resulting representations are combined with the residual connection and normalized,
\[
\tilde{\bH}^{(\ell)}
=
\mathrm{LayerNorm}\!\left(
\bH^{(\ell-1)}
+
\mathrm{MHA}\!\left(\bH^{(\ell-1)}\right)
\right).
\]
Let $\tilde{\mathbf{h}}^{(\ell)}_i\in\mathbb{R}^d$ denote the transpose of the $i$th row of $\tilde{\mathbf{H}}^{(\ell)}$.
A multi-layer perceptron is applied independently to each row of $\tilde{\mathbf{H}}^{(\ell)}$.
Let $d_{\mathrm{ff}}$ denote the hidden dimension of the multi-layer perceptron. For $i=0,1,\ldots,N$,
\[
\mathrm{MLP}\!\left(\tilde{\mathbf{h}}_i^{(\ell)}\right)
=
\mathbf{W}_{2}^{(\ell)}
\sigma\!\left(
\mathbf{W}_{1}^{(\ell)}
\tilde{\mathbf{h}}_i^{(\ell)}
+
\mathbf{b}_{1}^{(\ell)}
\right)
+
\mathbf{b}_{2}^{(\ell)},
\]
where $\mathbf{W}_{1}^{(\ell)}\in\mathbb{R}^{d_{\mathrm{ff}}\times d}$ and $\mathbf{W}_{2}^{(\ell)}\in\mathbb{R}^{d\times d_{\mathrm{ff}}}$ are learnable weight matrices.
The vectors $\mathbf{b}_{1}^{(\ell)}\in\mathbb{R}^{d_{\mathrm{ff}}}$ and $\mathbf{b}_{2}^{(\ell)}\in\mathbb{R}^{d}$ are learnable biases, and $\sigma(\cdot)$ denotes an elementwise nonlinear activation function.

After the final transformer layer, we extract the representation corresponding to the summary vector by taking the first row of $\bH^{(L)}$, denoted by $\bz_1\in\mathbb{R}^d$.
This vector aggregates information from all event representations through the self-attention mechanism.
Finally, we obtain the parameter estimate through the linear mapping
\[
\hat{\btheta}_1
=
\mathbf{W}_o\bz_1+\bb_o,
\]
where $\mathbf{W}_o\in\mathbb{R}^{d_{\btheta}\times d}$ and $\bb_o\in\mathbb{R}^{d_{\btheta}}$ are learnable parameters.

\section{Architecture of the Aggregation Network}
\label{sec:aggregator}

We describe the architecture of the aggregation network introduced in Section~\ref{sec:conquer}.
The aggregation network has a similar transformer-based architecture to the summary network described in Appendix~\ref{sec:summaryNetwork}, but it operates on block-level features rather than event-level inputs.
For block $j$, let $\mathbf{z}_j\in\mathbb{R}^d$ denote the block summary vector and let $\tilde{\mathbf{c}}_j\in\mathbb{R}^{p}$ denote the rescaled center coordinates of block $\mathcal{D}_j$.
We concatenate the block summary vector and the rescaled $p$-dimensional block-center coordinates to form
\[
\mathbf{u}_j
=
\begin{pmatrix}
\mathbf{z}_j\\
\tilde{\mathbf{c}}_j
\end{pmatrix}
\in\mathbb{R}^{d+p}.
\]
The resulting block feature vector is mapped to a $d$-dimensional representation through the linear projection
\[
\mathbf{r}_j
=
\mathbf{W}_{\mathrm{in}}\mathbf{u}_j
+
\mathbf{b}_{\mathrm{in}},
\]
where $\mathbf{W}_{\mathrm{in}}\in\mathbb{R}^{d\times(d+p)}$ and $\mathbf{b}_{\mathrm{in}}\in\mathbb{R}^{d}$ are learnable parameters.
Collecting over all $m$ blocks gives
\[
\mathbf{R}
=
\begin{pmatrix}
\mathbf{r}_1^\top\\
\vdots\\
\mathbf{r}_m^\top
\end{pmatrix}
\in\mathbb{R}^{m\times d}.
\]
This sequence is processed by the transformer encoder defined in \eqref{eq:transformer}, yielding $\mathbf{G}^{(L)}\in\mathbb{R}^{m\times d}$.
We denote the transpose of the $j$th row of $\mathbf{G}^{(L)}$ by $\mathbf{g}_j\in\mathbb{R}^{d}$, which is the contextualized representation of block $j$, obtained after information has been exchanged across blocks through self-attention.

To obtain a global representation, we apply attention pooling across blocks.
Let $\mathbf{q}\in\mathbb{R}^{d}$ denote a learnable query vector. For $j=1,\ldots,m$, we compute
\[
\alpha_j
=
\frac{
\exp(\mathbf{q}^\top\mathbf{g}_j/\sqrt d)
}{
\sum_{k=1}^{m}
\exp(\mathbf{q}^\top\mathbf{g}_k/\sqrt d)
},
\]
and define the global representation as
\[
\mathbf{g}
=
\sum_{j=1}^{m}
\alpha_j\mathbf{g}_j
\in\mathbb{R}^{d}.
\]

The parameter estimate is obtained by applying a multi-layer perceptron to the global representation.
Let $d_{\mathrm{mlp}}$ denote its hidden dimension. Specifically,
\[
\hat{\boldsymbol{\theta}}
=
\mathbf{W}_{o,2}
\sigma\!\left(
\mathbf{W}_{o,1}\mathbf{g}
+
\mathbf{b}_{o,1}
\right)
+
\mathbf{b}_{o,2},
\]
where $\mathbf{W}_{o,1}\in\mathbb{R}^{d_{\mathrm{mlp}}\times d}$ and $\mathbf{W}_{o,2}\in\mathbb{R}^{d_{\boldsymbol{\theta}}\times d_{\mathrm{mlp}}}$ are learnable weight matrices.
The vectors $\mathbf{b}_{o,1}\in\mathbb{R}^{d_{\mathrm{mlp}}}$ and $\mathbf{b}_{o,2}\in\mathbb{R}^{d_{\boldsymbol{\theta}}}$ are learnable biases.
For the quantile experiments, the single output MLP is replaced by the quantile-specific MLPs and monotone construction described in Section~\ref{sec:uncertainty}.

\section{Implementation Details}
\label{app:implementation}

Consider the architecture descriptions in Appendices~\ref{sec:summaryNetwork} and \ref{sec:aggregator}.
For the summary network, the input feature vectors are first projected into a $d$-dimensional embedding with $d=64$.
The resulting sequence of embeddings is then processed by the transformer encoder with $L=3$ layers, $H=4$ attention heads, and an MLP dimension of $d_{\mathrm{ff}}=128$.

For the aggregation network, the concatenated block feature vectors are first projected into a $d$-dimensional embedding with $d=64$.
The resulting sequence of block embeddings is then processed by a transformer encoder with $L=2$ layers, $H=4$ attention heads, and an MLP dimension of $d_{\mathrm{ff}}=128$.
Finally, the parameter estimate is obtained by applying a two-layer MLP to the global representation $\bg$, with hidden dimension $d_{\mathrm{mlp}}=128$.
For the posterior-quantile estimators used in the simulation experiments, each quantile-specific MLP has this two-layer architecture with hidden dimension $d_{\mathrm{mlp}}=128$.

For both networks, optimization is carried out using the Adam algorithm with a learning rate of $10^{-3}$. We use a mini-batch size of $128$ for the summary network and $256$ for the aggregation network.
All hidden layers use rectified linear unit (ReLU) activation functions.
The validation loss is monitored at each epoch, and early stopping is employed: model weights are saved whenever the validation loss improves, and training terminates if no improvement is observed for 10 consecutive epochs.

\section{Sensitivity to Block Length and Overlap}
\label{app:lgcp_block_sensitivity}

Using the LGCP example, we examine the sensitivity of our method to the block side length $\ell_b$ and the overlap length $\ell_o$ between adjacent blocks. We fix the full domain at $\mathcal D_S=[0,38]^2$. For an $n_b\times n_b$ grid of blocks, the settings satisfy
\begin{align*}
n_b\ell_b-(n_b-1)\ell_o=38.
\end{align*}
We consider three sets of block-partition settings. In the block-length experiment, we vary $\ell_b$ while keeping $\ell_o/\ell_b$ close to $0.3$. In the overlap experiment, we fix $\ell_b=10$ and vary $\ell_o/\ell_b$. We also consider a zero-overlap experiment in which $\ell_b$ is varied while $\ell_o=0$.

\begin{table}[!b]
\caption{Block-partition settings and median results over 30 training runs for the LGCP example. Block lengths are rounded to two decimal places. Summary RMSE is the joint RMSE of the block-level parameter estimates produced by the summary network; aggregation RMSEs are computed from the full-domain parameter estimates produced by the aggregation network. Computing time is the total training time for the summary and aggregation networks.}
\label{tab:lgcp_block_sensitivity_design}
\begin{center}
\small
\setlength{\tabcolsep}{5pt}
\begin{tabular}{lrrrrrrrrr}
\toprule
\multirow{2}{*}{Experiment}
& \multirow{2}{*}{$\ell_b$}
& \multirow{2}{*}{$\ell_o/\ell_b$}
& \multirow{2}{*}{Block grid}
& \multirow{2}{*}{\shortstack{Summary\\RMSE}}
& \multicolumn{4}{c}{Aggregation RMSE}
& \multirow{2}{*}{\shortstack{Computing\\time (min)}} \\
\cmidrule(lr){6-9}
 & & & & & Joint & $\gamma$ & $\kappa$ & $\rho$ & \\
\midrule
\multirow{5}{*}{Block length}
 & 3.00  & 0.31 & $18\times18$ & 0.963 & 0.436 & 0.346 & 0.347 & 0.575 & 47.2 \\
 & 5.00  & 0.27 & $10\times10$ & 0.932 & 0.438 & 0.351 & 0.351 & 0.574 & 11.4 \\
 & 7.50  & 0.32 & $7\times7$   & 0.890 & 0.438 & 0.360 & 0.357 & 0.565 & 12.8 \\
 & 10.00 & 0.30 & $5\times5$   & 0.850 & 0.446 & 0.369 & 0.363 & 0.573 & 20.1 \\
 & 12.50 & 0.32 & $4\times4$   & 0.809 & 0.452 & 0.377 & 0.369 & 0.577 & 27.6 \\
\midrule
\multirow{5}{*}{\shortstack[l]{Block length\\(zero overlap)}}
 & 1.19 & 0.00 & $32\times32$ & 0.990 & 0.452 & 0.366 & 0.369 & 0.582 & 432.1 \\
 & 1.90 & 0.00 & $20\times20$ & 0.980 & 0.443 & 0.356 & 0.358 & 0.576 & 68.8 \\
 & 4.75 & 0.00 & $8\times8$   & 0.936 & 0.449 & 0.356 & 0.357 & 0.594 & 12.6 \\
 & 6.33 & 0.00 & $6\times6$   & 0.910 & 0.449 & 0.360 & 0.359 & 0.588 & 13.0 \\
 & 9.50 & 0.00 & $4\times4$   & 0.857 & 0.451 & 0.364 & 0.363 & 0.585 & 19.6 \\
\midrule
\multirow{5}{*}{Block overlap}
 & 10.00 & 0.07 & $4\times4$   & 0.850 & 0.450 & 0.370 & 0.365 & 0.581 & 20.0 \\
 & 10.00 & 0.30 & $5\times5$   & 0.850 & 0.446 & 0.369 & 0.363 & 0.573 & 20.1 \\
 & 10.00 & 0.44 & $6\times6$   & 0.849 & 0.442 & 0.365 & 0.360 & 0.569 & 21.2 \\
 & 10.00 & 0.60 & $8\times8$   & 0.850 & 0.440 & 0.361 & 0.356 & 0.568 & 21.8 \\
 & 10.00 & 0.85 & $20\times20$ & 0.849 & 0.437 & 0.359 & 0.353 & 0.564 & 58.0 \\
\bottomrule
\end{tabular}
\end{center}
\end{table}

We use the same priors (i.e., $\gamma \sim \mathrm{Unif}(-4,-2)$, $\kappa \sim \mathrm{Unif}(0.1,5)$, and $\rho \sim \mathrm{Unif}(0.1,5)$) and network architectures as in Section~\ref{sec:sim_lgcp}.
For each block length, we generate 100,000 training and 2,000 validation point patterns on the corresponding block domain.
For aggregation-network training, all configurations use the same 100,000 training and 2,000 validation realizations on $\mathcal D_S=[0,38]^2$. We evaluate the estimators on a common set of $N=5{,}000$ test datasets. We repeat the complete training procedure 30 times. Thus, both the summary and aggregation networks are newly trained in each run.
For each block-partition setting and training run, we apply the summary network to every block in the common test set. For each parameter, we standardize every estimation error by dividing it by the corresponding prior standard deviation before computing the RMSEs. For the joint summary-network RMSE, we average the squared standardized errors between the true parameter values and the block-level parameter estimates produced by the summary network over blocks within each dataset and then over the test datasets and the three parameters, and take the square root. For the parameter-specific aggregation-network RMSEs, we average the squared standardized errors between the true parameter values and the full-domain parameter estimates produced by the aggregation network over the test datasets separately for each parameter; the joint aggregation-network RMSE additionally averages over the three parameters before taking the square root. Total computing time is the sum of the training times for the summary and aggregation networks.


In the block-length experiment, $\ell_o/\ell_b$ is kept close to $0.3$ while $\ell_b$ is varied. Table~\ref{tab:lgcp_block_sensitivity_design} shows that the summary-network RMSE decreases consistently as $\ell_b$ increases, indicating that the block-level parameter estimates produced by the summary network are more accurate for larger individual blocks. In contrast, the joint aggregation-network RMSE increases modestly from 0.436 at $\ell_b=3$ to 0.452 at $\ell_b=12.5$. Thus, the full-domain parameter estimates produced after aggregation tend to be slightly more accurate with smaller blocks, although the difference is small. The same trend is observed for $\gamma$ and $\kappa$, whereas the RMSE for $\rho$ is nonmonotone. Computing time is nonmonotone in block length: it decreases from 27.6 minutes at $\ell_b=12.5$ to 11.4 minutes at $\ell_b=5$, but then increases to 47.2 minutes at $\ell_b=3$. Larger blocks require more computation within the summary network, whereas very small blocks produce many summary vectors and therefore require more computation within the aggregation network.

To examine whether this pattern is related to overlap, we also vary the block length with zero overlap. The results are similar to those obtained with overlapping blocks. In this experiment, we further reduce the block length to $\ell_b=1.19$. At this value, the joint aggregation-network RMSE instead increases from 0.443 at $\ell_b=1.90$ to 0.452. At this resolution, many blocks contain only a few points or no points, leaving the summary network with little information for block-level estimation. Specifically, 82.8\% of the summary-network training point patterns are empty at $\ell_b=1.19$, compared with 66.6\% at $\ell_b=1.90$. This lack of information may increase the aggregation-network RMSE. This setting also requires substantially more computing time, taking 432.1 minutes.

Finally, in the block-overlap experiment, we fix $\ell_b=10$ and vary $\ell_o/\ell_b$. The aggregation-network RMSEs tend to decrease as $\ell_o/\ell_b$ increases, but the differences are small. For example, increasing $\ell_o/\ell_b$ from 0.07 to 0.85 reduces the joint aggregation-network RMSE from 0.450 to 0.437, while increasing the computing time from 20.0 to 58.0 minutes.

\begin{table}[!b]
\caption{Median aggregation-network RMSEs over 10 paired training runs when varying network capacity and aggregation training-set size at $\ell_b=7.6$ and $\ell_o=0$.}
\label{tab:lgcp_network_capacity_training}
\begin{center}
\small
\setlength{\tabcolsep}{5pt}
\begin{tabular}{lrrrrrr}
\toprule
\multirow{2}{*}{Condition}
& \multirow{2}{*}{\shortstack{Summary\\vector\\dimension}}
& \multirow{2}{*}{\shortstack{Aggregation\\training size}}
& \multicolumn{4}{c}{Aggregation RMSE} \\
\cmidrule(lr){4-7}
& & & Joint & $\gamma$ & $\kappa$ & $\rho$ \\
\midrule
Baseline & 64 & 100,000 & 0.457 & 0.369 & 0.364 & 0.597 \\
Expanded summary network & 128 & 100,000 & 0.457 & 0.371 & 0.363 & 0.598 \\
Expanded aggregation network & 64 & 100,000 & 0.459 & 0.371 & 0.367 & 0.599 \\
More aggregation data & 64 & 200,000 & 0.454 & 0.365 & 0.363 & 0.594 \\
Expanded aggregation $+$ more data & 64 & 200,000 & 0.455 & 0.368 & 0.363 & 0.594 \\
\bottomrule
\end{tabular}
\end{center}
\end{table}

To assess whether the slightly poorer results for larger blocks reflect limited network capacity or insufficient aggregation training data, we conduct four additional experiments at $\ell_b=7.6$ and $\ell_o=0$. Relative to the baseline, we expand the summary network and increase its summary-vector dimension from 64 to 128, expand the aggregation network, double the aggregation training set from 100,000 to 200,000 realizations, and combine the latter two changes. In the expanded summary network, the representation dimension, number of attention heads, and feed-forward dimension increase from 64, 4, and 128 to 128, 8, and 256, respectively, while the number of transformer layers remains three. The expanded aggregation network increases the representation dimension from 64 to 128, the number of attention heads from 4 to 8, the number of transformer layers from 2 to 3, and both the feed-forward and output-head dimensions from 128 to 256. All conditions use the same validation and test data and are compared over 10 paired training runs; the two larger-data conditions append 100,000 training realizations to the baseline training set. The RMSEs are computed using the same standardization as in Table~\ref{tab:lgcp_block_sensitivity_design}.

Table~\ref{tab:lgcp_network_capacity_training} shows that expanding the summary network and increasing its summary-vector dimension from 64 to 128 produces no consistent change in the aggregation-network RMSEs. Expanding the aggregation network alone also does not improve accuracy. Doubling the aggregation training set yields small reductions in the joint and parameter-specific aggregation-network RMSEs, and combining more data with the expanded aggregation network gives similar improvements. These results provide little evidence that limited network capacity drives the observed pattern across block lengths; additional aggregation training data improves accuracy only modestly.

One possible explanation for this behavior is the number and spatial resolution of the summary vectors supplied to the aggregation network. Smaller blocks yield less accurate individual summaries, but together these summaries provide a finer-grained representation of the full spatial pattern. In contrast, a summary vector from a larger block must compress information over a wider region and may therefore obscure some local spatial variation. The aggregation network may thus compensate for weaker individual summaries by combining many localized summaries, leading to similar or slightly better aggregation accuracy with smaller blocks. However, this pattern does not extend to extremely small blocks: when many blocks contain only a few or no points, their summaries provide too little information and aggregation accuracy deteriorates. Overall, our method is relatively insensitive to block length and overlap over the ranges considered, except at the smallest block length.

\section{Proofs and Derivations}
\label{sup:sec:information}

Throughout this appendix, the fitted maps are treated as fixed, and all relevant maps are assumed to be measurable.

\subsection{Proof of Proposition~\ref{prop:general_information}}
\label{sup:sec:bayes_risk_monotonicity}

Let $\mathcal{D}_Z$ and $\mathcal{D}_U$ denote the classes of all measurable estimators based on $(\bZ_{\mathcal C},\mathcal C)$ and $(\bU_{\mathcal C},\mathcal C)$, respectively, taking values in $\bTheta$.
Let $\mathcal R_Z(L)$ and $\mathcal R_U(L)$ denote the sets of risks under the loss $L(\cdot,\cdot)$ attainable by estimators in $\mathcal D_Z$ and $\mathcal D_U$, respectively.
Since $\bU_{\mathcal C}$ is obtained by applying $T_{\bdelta^*}(\cdot)$ blockwise to $\bZ_{\mathcal C}$, every measurable estimator of the form $d_U(\bU_{\mathcal C},\mathcal C)$ can be written as $d_Z(\bZ_{\mathcal C},\mathcal C)$ for some $d_Z\in\mathcal D_Z$. Hence,
\[
\mathcal R_U(L)\subseteq\mathcal R_Z(L).
\]
Taking infima over the two sets, the corresponding Bayes risks satisfy
\[
R_Z^*(L)\le R_U^*(L),
\]
which proves Proposition~\ref{prop:general_information}.
\hfill\BlackBox

\subsection{Proof of Corollary~\ref{cor:squared_information}}
\label{sup:sec:squared_information_gap}

Define
\[
\mathcal{G}
=
\sigma(\bZ_{\mathcal C},\mathcal C),
\qquad
\mathcal{H}
=
\sigma(\bU_{\mathcal C},\mathcal C),
\]
where $\sigma(\cdot)$ denotes the $\sigma$-field generated by its arguments.
Since $\bU_{\mathcal C}$ is obtained from $\bZ_{\mathcal C}$ by applying the measurable map $T_{\bdelta^*}(\cdot)$ to each blockwise summary, we have $\mathcal{H}\subseteq\mathcal{G}$.

Let
\[
\mathbf{m}_Z
=
\mathbb{E}(\btheta\mid\mathcal{G}),
\qquad
\mathbf{m}_U
=
\mathbb{E}(\btheta\mid\mathcal{H}).
\]
Under squared-error loss, $\mathbf{m}_Z$ and $\mathbf{m}_U$ are the Bayes estimators based on $\mathcal{G}$ and $\mathcal{H}$, respectively. The corresponding Bayes risks are
\[
R_Z^*
=
\mathbb{E}\|\btheta-\mathbf{m}_Z\|^2,
\qquad
R_U^*
=
\mathbb{E}\|\btheta-\mathbf{m}_U\|^2.
\]
To compare these risks, write
\[
\btheta-\mathbf{m}_U
=
(\btheta-\mathbf{m}_Z)
+
(\mathbf{m}_Z-\mathbf{m}_U).
\]
Taking squared norms and expectations gives
\[
R_U^*
=
R_Z^*
+
\mathbb{E}\|\mathbf{m}_Z-\mathbf{m}_U\|^2
+
2\mathbb{E}\left[
(\btheta-\mathbf{m}_Z)^\top
(\mathbf{m}_Z-\mathbf{m}_U)
\right].
\]
Because $\mathcal{H}\subseteq\mathcal{G}$, both $\mathbf{m}_Z$ and $\mathbf{m}_U$ are $\mathcal{G}$-measurable. Hence, by the tower property,
\[
\begin{aligned}
\mathbb{E}\left[
(\btheta-\mathbf{m}_Z)^\top
(\mathbf{m}_Z-\mathbf{m}_U)
\right]
&=
\mathbb{E}\left[
\mathbb{E}(\btheta-\mathbf{m}_Z\mid\mathcal{G})^\top
(\mathbf{m}_Z-\mathbf{m}_U)
\right]
\\
&=0,
\end{aligned}
\]
since $\mathbb{E}(\btheta-\mathbf{m}_Z\mid\mathcal{G})=\bzero$. Therefore,
\[
R_U^*-R_Z^*
=
\mathbb{E}\|\mathbf{m}_Z-\mathbf{m}_U\|^2
\geq 0.
\]
Equality holds if and only if $\mathbf{m}_Z=\mathbf{m}_U$ almost surely.
\hfill\BlackBox

\subsection{Derivation for Remark~\ref{rem:learned_risk}}
\label{sup:sec:fitted_rules}

Consider a nonnegative measurable loss $L(\cdot,\cdot)$ for which all risks below are finite. Let the fitted rules be as defined in Remark~\ref{rem:learned_risk}:
\begin{align*}
\hat{\btheta}_Z
&=
g_{\bpsi^*}(\bZ_{\mathcal C}), \\
\hat{\btheta}_U
&=
a^*(\bU_{\mathcal C}).
\end{align*}

For any fitted rule $\hat{\btheta}$, write
\[
R(\hat{\btheta};L)
=
\mathbb E\{L(\btheta,\hat{\btheta})\}.
\]
Define the excess risks relative to the respective Bayes rules by
\begin{align*}
\mathcal E_Z(L)
&=
R(\hat{\btheta}_Z;L)-R_Z^*(L),\\
\mathcal E_U(L)
&=
R(\hat{\btheta}_U;L)-R_U^*(L).
\end{align*}
Both excess risks are nonnegative because each fitted rule belongs to the corresponding class of measurable estimators used to define the Bayes risk. By their definitions,
\begin{align*}
R(\hat{\btheta}_Z;L)
&=
R_Z^*(L)+\mathcal E_Z(L),\\
R(\hat{\btheta}_U;L)
&=
R_U^*(L)+\mathcal E_U(L).
\end{align*}
Subtracting the first identity from the second gives
\begin{align*}
R(\hat{\btheta}_U;L)-R(\hat{\btheta}_Z;L)
=
\{R_U^*(L)-R_Z^*(L)\}
+
\{\mathcal E_U(L)-\mathcal E_Z(L)\}.
\end{align*}
The first term is nonnegative by Proposition~\ref{prop:general_information}. Moreover,
\[
R(\hat{\btheta}_Z;L)
<
R(\hat{\btheta}_U;L)
\quad\Longleftrightarrow\quad
\mathcal E_Z(L)-\mathcal E_U(L)
<
R_U^*(L)-R_Z^*(L),
\]
which is the condition stated in Remark~\ref{rem:learned_risk}.
\par

\clearpage

\section{Predicted Intensity Surfaces for Whale Call Data}
\label{app:whale-intensity}

\begin{figure}[htbp]
\centering
\includegraphics[width=0.7\linewidth]{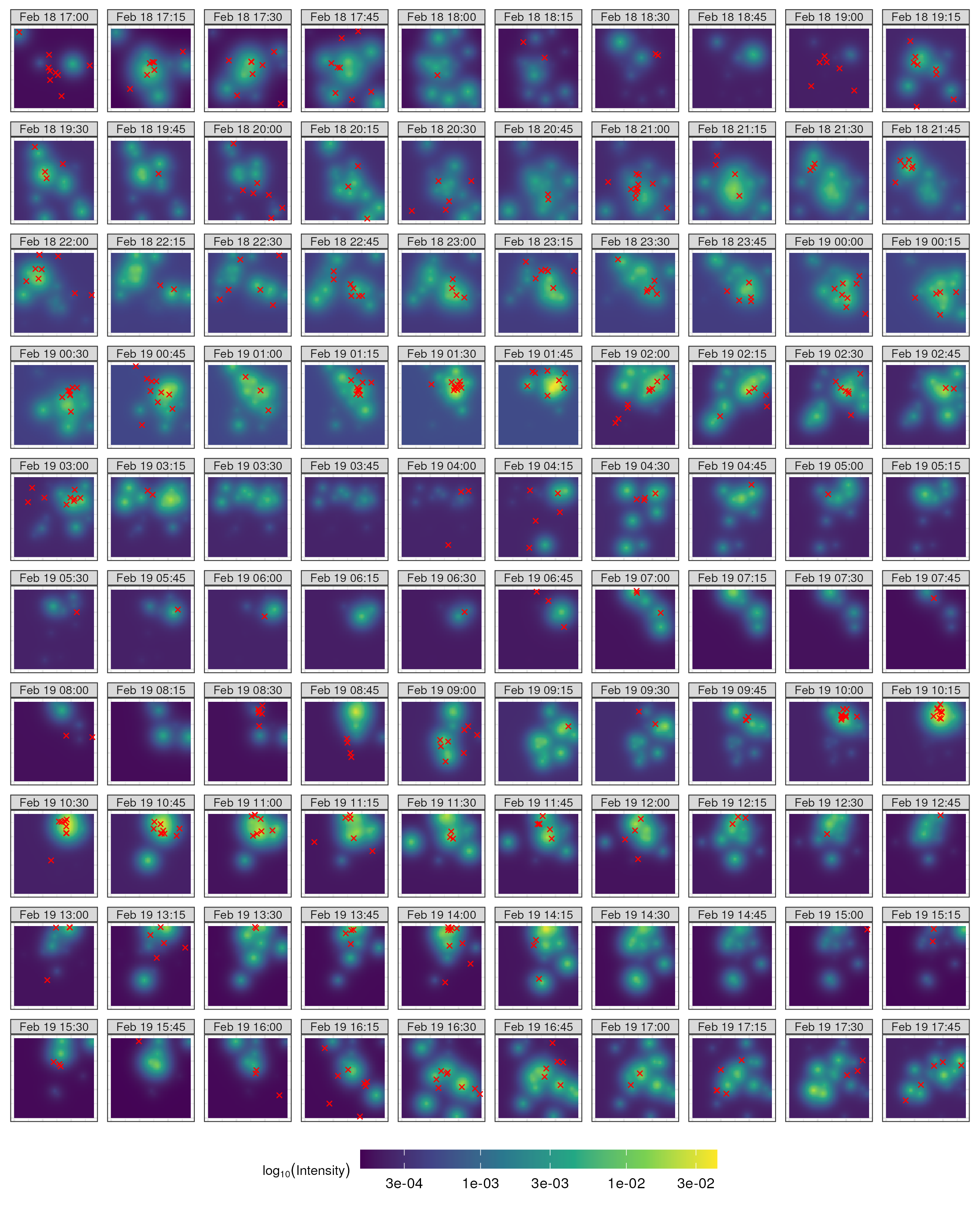}
\caption{Predicted 15-minute-ahead intensity surfaces and corresponding observed data (red crosses) over consecutive intervals for whale call data fitted to the spatiotemporal Hawkes process model.}
\label{sup:fig:real9_auto3_pred}
\end{figure}

\clearpage

\vskip 0.2in
\bibliographystyle{apalike}
\bibliography{refs}

\end{document}